\documentclass[aps,prd,superscriptaddress,amsmath,groupedaddress,       
         twocolumn,showpacs]{revtex4}
\usepackage{amssymb,graphicx}
\usepackage{dcolumn}

\newcommand{\msb}{{\rm\overline{MS}}}
\newcommand{\almsb}{\alpha_\msb}
\newcommand{\eq}[1]{Eq.~(\ref{#1})}
\newcommand{\order}{{\cal O}}
\newcommand{\xv}{{\bf x}}
\newcommand{\psib}{\overline{\psi}}
\newcommand{\g}{\gamma}

\newcommand{\zmc}{1.507(7)}
\newcommand{\mcmu}{0.991(5)}
\newcommand{\mcmc}{1.273(6)}
\newcommand{\mc}{0.986(6)}
\newcommand{\zmb}{1.296(8)}
\newcommand{\mbmu}{3.622(22)}
\newcommand{\mbmb}{4.164(23)}
\newcommand{\mb}{3.617(25)}
\newcommand{\mbmc}{4.53(4)}
\newcommand{\alphaMZ}{0.1183(7)}
\newcommand{\alphafit}{0.2034(21)}

\begin{document}
	\title{High-Precision $c$ and $b$ Masses, and QCD Coupling\\
	        from Current-Current Correlators in Lattice and Continuum QCD}
  	\author{C.\ McNeile}
   \thanks{Current address: Dept.\ of Theoretical Physics, Univ.\ of
    Wuppertal, Wuppertal 42199, Germany  }
  	\affiliation{Department of Physics and Astronomy, University of 
Glasgow,
        	 	Glasgow G12 8QQ, UK}
	\author{C.\ T.\ H.\ Davies}
	\affiliation{Department of Physics and Astronomy, University of 
Glasgow,
	 	Glasgow G12 8QQ, UK}
	\author{E.\ Follana}
	\affiliation{Departamento de F\'{\i}sica Te\'{o}rica, Universidad de
	 Zaragoza, E-50009 Zaragoza, Spain}
	\author{K.\ Hornbostel}
	\affiliation{Southern Methodist University, Dallas, Texas 75275, USA}
	\author{G.\ P.\ Lepage}
	\email{g.p.lepage@cornell.edu}
	\affiliation{Laboratory for Elementary-Particle Physics,
		Cornell University, Ithaca, NY 14853, USA}
	\collaboration{HPQCD Collaboration}
	\noaffiliation
	\date{March 26, 2010}

\begin{abstract}
   We extend our earlier lattice-QCD analysis of heavy-quark 
correlators to smaller lattice spacings and larger masses to obtain new 
values for the $c$~mass and QCD coupling, and, for the first time, 
values for the $b$~mass: $m_c(3\,\mathrm{GeV},n_f\!=\!4)\!=\!\mc$\,GeV, 
$\alpha_\msb(M_Z,n_f\!=\!5)\!=\!\alphaMZ$, and 
$m_b(10\,\mathrm{GeV},n_f\!=\!5)\!=\!\mb$\,GeV. These are among the 
most accurate determinations by any method. We check our results using 
a  nonperturbative determination of the mass 
ratio~$m_b(\mu,n_f)/m_c(\mu,n_f)$; the two methods agree to within our 
1\%~errors and taken together imply~$m_b/m_c\!=\!4.51(4)$. We also 
update our previous analysis of $\alpha_\msb$ from Wilson loops to 
account for revised values for $r_1$ and $r_1/a$, finding a new value 
$\alpha_\msb(M_Z,n_f\!=\!5)\!=\!0.1184(6)$; and we update our recent 
values for light-quark masses from the ratio~$m_c/m_s$. Finally, in the 
Appendix, we derive a procedure for simplifying and accelerating 
complicated least-squares fits.
\end{abstract}

\maketitle

\section{Introduction}
Precise values for the QCD coupling $\alpha_\msb$ and the quark masses 
are important for high-precision tests of the Standard Model of 
particle physics. In a recent paper we showed how to use realistic 
lattice QCD simulations to extract both the coupling and the charm 
quark's mass~$m_c$ from zero-momentum moments of correlators built from 
the $c$~quark's (UV cutoff-independent) pseudoscalar density 
operator~$m_c\psib_c\gamma_5\psi_c$~\cite{mcjj}. In this paper we 
refine our previous analysis and extend it to include other quark 
masses, up to and including the $b$-quark mass. As a result our 
coupling constant and mass determinations from these correlators are 
among the most accurate by any method.

Low moments of heavy-quark correlators are perturbative and several are 
now known through $\order(\alpha_s^3)$ in perturbation theory (that is, 
four-loop 
order)~\cite{Chetyrkin:2006xg,Boughezal:2006px,Maier:2008he,Maier:2009fz,Kiyo:2009gb}.
Moments of correlators built from the electromagnetic currents can be 
estimated nonperturbatively, using dispersion relations, from 
experimental data for the electron-positron annihilation cross section, 
$\sigma(e^+ e^-\!\to\!\gamma^*\!\to\!X)$. Accurate values for both the 
$c$~and $b$~masses can be obtained by comparing these perturbative and 
nonperturbative determinations of the moments (for a recent discussion 
see~\cite{Chetyrkin:2009fv}).

In our earlier paper we showed that heavy-quark correlator moments are 
easily and accurately computed nonperturbatively using lattice QCD 
simulations, in place of experimental data, provided: 1) the 
electromagnetic current is replaced by the pseudoscalar density 
multiplied by the bare quark mass; 2) the discretization of the quark 
action has a partially conserved axial vector current (PCAC); and 3) 
the discretization remains accurate when applied to heavy quarks. In 
our simulations we use the HISQ discretization of the quark action, 
which is a highly corrected version of the standard staggered-quark 
action~\cite{Follana:2006rc}. It has a chiral symmetry (PCAC) and has 
been used in a wide variety of accurate simulations involving 
$c$~quarks~\cite{Follana:2006rc,Follana:2007uv,Davies:2009tsa,Davies:2009ih,Gregory:2009hq}.

Here we show that the HISQ action can be pushed to still higher 
masses\,---\,indeed, very close to the $b$~mass\,---\,on new lattices, 
from the MILC collaboration~\cite{MILC}, with the smallest lattice 
spacings available today (0.044\,fm). Currently most high-precision 
lattice work on $b$~physics relies upon nonrelativistic effective field 
theories, like 
NRQCD~\cite{Davies:2009tsa,Gregory:2009hq,Gamiz:2009ku,Gray:2005ur}. In 
this paper we show how to obtain accurate $b$~physics using the fully 
relativistic HISQ action on these new lattices.

In what follows, we first review how the QCD coupling and quark masses 
are extracted from heavy-quark correlators, in Section~II. Then in 
Section~III we describe our lattice QCD simulations and discuss in 
detail the chief systematic errors in our simulation results. In 
Section~IV we describe our fitting procedure and the results of our 
analysis of the heavy-quark correlators. We check our calculation using 
a different, nonperturbative method to determine~$m_b/m_c$ in 
Section~V. We then, in Section~VI, update our previous calculation of 
the QCD coupling from Wilson loops to compare with our new result from 
the correlators. We summarize our findings in Section~VII and compare 
our results with work by others. There we also update our recent 
calculations of the light-quark masses from the $c$~mass. In the 
Appendix we present a powerful simplification for complicated 
least-squares fits that can greatly reduce the computing required for 
fits. We use this technique in dealing with finite-$a$ errors in our 
analysis.

\section{Heavy-Quark Correlator Moments}
Following our earlier paper~\cite{mcjj}, we focus on correlators formed 
from the pseudoscalar density of a heavy quark, $j_5 \!=\! 
\psib_h\g_5\psi_h$:
\begin{equation} \label{G}
   G(t) = a^6 \sum_\xv (am_{0h})^2 \langle 0 | j_5(\xv,t) j_5(0,0)
         |0\rangle
\end{equation}
where $m_{0h}$ is the heavy quark's bare mass (from the lattice QCD 
lagrangian), $t$ is euclidean and periodic with period~$T$, and the sum 
over spatial positions~$\xv$ sets the total three momentum to zero. In 
our earlier paper we examined only $c$~quarks; here we will consider a 
range of masses between the $c$~and $b$~masses. While we have written 
this formula for use with the lattice regulator, it is important to 
note that the correlator is UV-finite because we include the factors of 
$am_{0h}$. Consequently lattice and continuum $G(t)$s are equal when 
$t\ne0$ up to $\order((am_h)^m)$ corrections, which vanish in the 
continuum limit.

The moments of $G(t)$ are particularly simple to analyze:
\begin{equation}
   G_n \equiv \sum_t (t/a)^n G(t)
\end{equation}
where, on our periodic lattice,
\begin{equation}
   t/a \in \{0,1,2\ldots T/2a-1,0,-T/2a+1\ldots -2,-1\}.
\end{equation}
Low moments emphasize small $t$s and so are perturbative; and moments 
with $n\ge4$ are UV-cutoff independent. Therefore
\begin{equation} \label{gn-def}
   G_n = \frac{g_n(\almsb(\mu),\mu/m_h)}{(am_h(\mu))^{n-4}} +
    \order((am_h)^m)
\end{equation}
for small $n\ge4$, where $m_h(\mu)$ is the heavy quark's $\msb$ mass at 
scale $\mu$, and the dimensionless factor $g_n$ can be computed using 
continuum perturbation theory.

Again following our previous paper, we introduce reduced moments to 
suppress both lattice artifacts and tuning errors in the heavy quark's 
mass~\cite{mpole}:
\begin{equation}\label{Rn-def}
    R_n \equiv \left\{
    \begin{aligned}
        & G_4/G^{(0)}_4 & & \text{for $n=4$,} \\
        &\frac{a m_{\eta_h}}{2a m_{0h}}
            \left(G_n/G^{(0)}_n\right)^{1/(n-4)}
        & & \text{for $n\ge6$,}
    \end{aligned}\right.
\end{equation}
where $G^{(0)}_n$ is the moment in lowest-order, weak-coupling 
perturbation theory, using the lattice regulator, and $m_{\eta_h}$ is 
the (nonperturbative) mass of the pseudo-Goldstone 
$h\overline{h}$~boson.
The reduced moments can again be written in terms of continuum 
quantities:
\begin{equation}\label{Rn-eqn}
    R_n \equiv \left\{
	\begin{aligned}
		& r_4(\alpha_\msb,\mu/m_h) & & \text{for $n=4$,} \\
		& z(\mu/m_h,m_{\eta_h})\,r_n(\alpha_\msb,\mu/m_h)
		& & \text{for $n\ge6$,}
	\end{aligned}\right.
   \end{equation}
up to $\order((am_h)^m\alpha_s)$ corrections, where
\begin{equation} \label{z-def}
   z(\mu/m_h,m_{\eta_h}) \equiv \frac{m_{\eta_h}}{2m_h(\mu)},
\end{equation}
and $r_n$ is obtained from $g_n$ (\eq{gn-def}) and its value, 
$g_n^{(0)}$, in lowest-order continuum perturbation theory:
\begin{equation}\label{rn-def}
    r_n = \begin{cases} g_4/g_4^{(0)} & \text{for $n=4$,}\\
                \left(g_n/g_n^{(0)}\right)^{1/(n-4)} & \text{for 
$n\ge6$.}
        \end{cases}
\end{equation}

Our strategy for extracting quark masses and the QCD coupling relies 
upon lattice simulations to determine nonperturbative values for the 
$R_n$, using simulation results for $am_{\eta_h}/am_{0h}$. We then 
compare this simulation ``data'' to the continuum perturbation theory 
formulas (\eq{Rn-eqn}). That is, we find values for $\almsb(\mu)$ and 
$z(\mu/m_h,m_{\eta_h})$ that make lattice and continuum results agree 
for small $n\ge4$. The function $z(\mu/m_h,m_{\eta_h})$ can then be 
combined with experimental results for $m_{\eta_c}$ and $m_{\eta_b}$ to 
obtain masses for the $c$~and $b$~quarks:
\begin{equation}\label{m-determination}
   m_c(\mu) = \frac{m_{\eta_c}^{\mathrm{exp}}}   
{2z(\mu/m_c,m_{\eta_c}^\mathrm{exp})}
   \quad\quad
   m_b(\mu) = \frac{m_{\eta_b}^{\mathrm{exp}}}   
{2z(\mu/m_b,m_{\eta_b}^\mathrm{exp})}
\end{equation}

Parameter~$\mu$ sets the scale for~$\alpha_\msb$ in the perturbative 
expansions of the~$r_n$. An obvious choice for this parameter is 
$\mu\!=\!m_h$ since the quark mass, together with~$n$, sets the 
momentum scale in our correlators. As noted in our previous paper, 
however, perturbation theory is somewhat more convergent if we use 
larger $\mu$s in the $c$-quark case. Consequently here we take 
$\mu/m_h\!=\!3$, which is approximately what we did in our previous 
paper.

The mass and coupling determinations were done separately in our 
previous paper. Here we extract them simultaneously, to guarantee 
consistency between results. Also in our previous paper we considered 
only heavy-quark masses near the $c$~mass. Here we explore a variety of 
masses ranging from just below the $c$~mass to just below the $b$~mass. 
This allows us to obtain a value for $b$-quark's mass.

\section{Lattice QCD Simulations}

\begin{table}
   \caption{Parameter sets used to generate the gluon configurations
   analyzed in this paper. The lattice spacing is specified in terms
   of the static-quark potential parameter $r_1\!=\!0.3133(23)$\,fm; 
values
   for $r_1/a$ are from~\cite{MILC}.
   The bare quark masses are for the ASQTAD formalism and $u_0$
   is the fourth root of the plaquette. The spatial ($L$) and temporal
   ($T$) lengths of the lattices are also listed, as are the number
   of gluon configurations ($N_\mathrm{cf}$) and the number of time 
sources
   ($N_\mathrm{ts}$) per configuration used in each case.
   Sets with similar lattice spacings are
   grouped.}
   \label{tab:cfg}
   \begin{ruledtabular}
      \begin{tabular}{cccccccc}
      Set & $r_1/a$ & $au_0m_{0u/d}$ & $au_0m_{0s}$ & $u_0$ & $L/a$ & 
$T/a$
      & $N_\mathrm{cf}\times N_\mathrm{ts}$
      \\ \hline
      1 & 2.152(5) & 0.0097 & 0.0484 & 0.860 & 16 & 48 & $631\times2$ 
\\
      2 & 2.138(4) & 0.0194 & 0.0484 & 0.861 & 16 & 48 & $631\times2$ 
\\
       \hline
      3 & 2.647(3) & 0.005 & 0.05 & 0.868 & 24 & 64 & $678\times2$ \\
      4 & 2.618(3) & 0.01 & 0.05 & 0.868 & 20 & 64 & $595\times2$ \\
      5 & 2.618(3) & 0.01 & 0.05 & 0.868 & 28 & 64 & $269\times2$ 
\\ \hline
      6 & 3.699(3) & 0.0062 & 0.031 & 0.878 & 28 & 96 & $566\times4$\\
      7 & 3.712(4) & 0.0124 & 0.031 & 0.879 & 28 & 96 & $265\times4$ \\
       \hline
      8 & 5.296(7) & 0.0036 & 0.018 & 0.888 & 48 & 144 & $201\times2$ 
\\
       \hline
      9 & 7.115(20) & 0.0028 & 0.014 & 0.895 & 64 & 192  & 
$208\times2$\\
      \end{tabular}
   \end{ruledtabular}
\end{table}

\subsection{Simulation Results}
The gluon-configuration sets we use were created by the MILC 
collaboration. The relevant simulation parameters are listed in 
Table~\ref{tab:cfg}.

\begin{table*}
   \caption{Results for the reduced moments~$R_n$ and 
pseudoscalar-meson mass~$am_{\eta_h}$ obtained from ($n_f\!=\!3$) 
simulations using different bare heavy-quark (HISQ) masses $am_{0h}$ 
and gluon configuration sets (see Table~\ref{tab:cfg}). The errors 
listed here are statistical errors from the Monte Carlo simulation. 
Results where $am_{\eta_h}\!>\!1.95$ are omitted from our final 
analysis, as are $R_n$s with $n\!>\!10$.}
   \label{tab:Rn-MC}
\begin{ruledtabular}
   \begin{tabular}{ccc|cccccccc}
   Set & $am_{0h}$ & $am_{\eta_h}$ & $R_4$ & $R_6$ & $R_8$ & $R_{10}$ & 
$R_{12}$ & $R_{14}$ & $R_{16}$ & $R_{18}$  \\ \hline
   1 &  0.660 & 1.9202(1) & 1.2132(3) & 1.5364(3) & 1.4151(2) & 
1.3476(1) & 1.3001(1) & 1.2649(1) & 1.2378(1) & 1.2164(1) \\
     &  0.810 & 2.1938(1) & 1.1643(2) & 1.4427(2) & 1.3619(1) & 
1.3148(1) & 1.2780(1) & 1.2481(1) & 1.2238(1) & 1.2039(1) \\
     &  0.825 & 2.2202(1) & 1.1604(2) & 1.4339(2) & 1.3563(1) & 
1.3111(1) & 1.2754(1) & 1.2462(1) & 1.2222(1) & 1.2025(1) \\
   2 &  0.825 & 2.2196(1) & 1.1591(2) & 1.4327(2) & 1.3556(1) & 
1.3106(1) & 1.2751(1) & 1.2459(1) & 1.2221(1) & 1.2024(1) \\
   3 &  0.650 & 1.8458(1) & 1.1809(2) & 1.4805(2) & 1.3755(1) & 
1.3160(1) & 1.2740(1) & 1.2429(1) & 1.2190(1) & 1.2000(1) \\
   4 &  0.440 & 1.4241(1) & 1.2752(4) & 1.6144(4) & 1.4397(2) & 
1.3561(2) & 1.3041(1) & 1.2678(1) & 1.2408(1) & 1.2200(1) \\
     &  0.630 & 1.8085(1) & 1.1881(3) & 1.4935(2) & 1.3826(1) & 
1.3205(1) & 1.2773(1) & 1.2456(1) & 1.2214(1) & 1.2021(1) \\
     &  0.660 & 1.8667(1) & 1.1782(2) & 1.4764(2) & 1.3738(1) & 
1.3152(1) & 1.2736(1) & 1.2426(1) & 1.2187(1) & 1.1997(1) \\
     &  0.720 & 1.9811(1) & 1.1605(2) & 1.4435(2) & 1.3559(1) & 
1.3044(1) & 1.2662(1) & 1.2367(1) & 1.2136(1) & 1.1950(1) \\
     &  0.850 & 2.2194(1) & 1.1301(2) & 1.3763(1) & 1.3145(1) & 
1.2774(1) & 1.2473(1) & 1.2221(1) & 1.2012(1) & 1.1839(1) \\
   5 &  0.630 & 1.8086(1) & 1.1882(1) & 1.4936(1) & 1.3826(1) & 
1.3205(1) & 1.2774(1) & 1.2457(1) & 1.2214(0) & 1.2022(0) \\
   6 &  0.300 & 1.0314(1) & 1.2930(3) & 1.6061(3) & 1.4249(2) & 
1.3444(1) & 1.2953(1) & 1.2610(1) & 1.2353(1) & 1.2153(1) \\
     &  0.413 & 1.2806(1) & 1.2224(2) & 1.5216(2) & 1.3796(1) & 
1.3115(1) & 1.2689(1) & 1.2390(1) & 1.2164(1) & 1.1985(1) \\
     &  0.430 & 1.3169(1) & 1.2145(2) & 1.5113(2) & 1.3743(1) & 
1.3076(1) & 1.2658(1) & 1.2363(1) & 1.2141(1) & 1.1964(1) \\
     &  0.440 & 1.3382(1) & 1.2100(2) & 1.5054(2) & 1.3712(1) & 
1.3054(1) & 1.2640(1) & 1.2348(1) & 1.2127(1) & 1.1952(1) \\
     &  0.450 & 1.3593(1) & 1.2057(2) & 1.4996(2) & 1.3683(1) & 
1.3033(1) & 1.2623(1) & 1.2333(1) & 1.2114(1) & 1.1941(1) \\
     &  0.700 & 1.8654(1) & 1.1301(1) & 1.3782(1) & 1.3053(1) & 
1.2616(1) & 1.2294(1) & 1.2048(0) & 1.1857(0) & 1.1705(0) \\
     &  0.850 & 2.1498(1) & 1.1026(1) & 1.3163(1) & 1.2671(1) & 
1.2366(0) & 1.2114(0) & 1.1903(0) & 1.1729(0) & 1.1584(0) \\
   7 &  0.427 & 1.3074(1) & 1.2131(3) & 1.5091(3) & 1.3729(2) & 
1.3066(1) & 1.2651(1) & 1.2358(1) & 1.2137(1) & 1.1961(1) \\
   8 &  0.273 & 0.8993(3) & 1.2454(8) & 1.5234(9) & 1.3739(7) & 
1.3069(6) & 1.2657(6) & 1.2366(6) & 1.2145(6) & 1.1969(5) \\
     &  0.280 & 0.9154(2) & 1.2403(5) & 1.5175(5) & 1.3706(3) & 
1.3045(3) & 1.2638(2) & 1.2350(2) & 1.2132(2) & 1.1958(2) \\
     &  0.564 & 1.5254(1) & 1.1324(2) & 1.3674(2) & 1.2857(2) & 
1.2405(1) & 1.2102(1) & 1.1885(1) & 1.1719(1) & 1.1587(1) \\
     &  0.705 & 1.8084(1) & 1.1043(2) & 1.3156(2) & 1.2574(1) & 
1.2217(1) & 1.1952(1) & 1.1750(1) & 1.1593(1) & 1.1467(1) \\
     &  0.760 & 1.9157(1) & 1.0955(2) & 1.2965(2) & 1.2460(1) & 
1.2142(1) & 1.1895(1) & 1.1701(1) & 1.1547(1) & 1.1423(1) \\
     &  0.850 & 2.0875(1) & 1.0831(2) & 1.2666(1) & 1.2266(1) & 
1.2010(1) & 1.1799(1) & 1.1621(1) & 1.1474(1) & 1.1353(1) \\
   9 &  0.195 & 0.6710(2) & 1.2583(5) & 1.5243(5) & 1.3733(4) & 
1.3066(3) & 1.2655(3) & 1.2364(2) & 1.2144(2) & 1.1968(2) \\
     &  0.400 & 1.1325(2) & 1.1532(3) & 1.3800(3) & 1.2838(2) & 
1.2370(2) & 1.2077(2) & 1.1869(2) & 1.1710(2) & 1.1583(2) \\
     &  0.500 & 1.3446(2) & 1.1267(2) & 1.3410(2) & 1.2616(1) & 
1.2198(1) & 1.1927(1) & 1.1734(1) & 1.1588(1) & 1.1471(1) \\
     &  0.700 & 1.7518(1) & 1.0900(1) & 1.2765(1) & 1.2261(1) & 
1.1949(1) & 1.1718(1) & 1.1542(1) & 1.1407(1) & 1.1299(1) \\
     &  0.850 & 2.0428(1) & 1.0712(1) & 1.2327(1) & 1.1983(1) & 
1.1760(1) & 1.1574(1) & 1.1418(1) & 1.1290(1) & 1.1185(1) \\
   \end{tabular}
\end{ruledtabular}
\end{table*}

Given a lattice spacing, the QCD action is specified completely by the 
values of the bare coupling constant and the bare quark masses. In our 
analyses we set the $u$~and $d$~quark masses equal; this approximation 
results in negligible errors ($\ll 1\%$) for the quantities studied in 
this paper. It is too costly to simulate QCD at the correct value for 
the $u/d$ mass; we typically use masses that are 2--5~times too large 
and extrapolate to values that give the correct mass for the 
$\pi^0$-meson. We tune the strange quark mass to give the correct mass 
for the (fictitious) $\eta_s$~meson~\cite{Davies:2009tsa}. The $c$~and 
$b$~masses are tuned to give correct masses for the $\eta_c$~and 
$\eta_b$~mesons, respectively.

It is convenient in QCD simulations to specify a value for the bare 
coupling constant and then extract the value of the lattice spacing 
from the simulation. We set the lattice spacing using MILC results for 
$r_1/a$, computed from the heavy-quark potential, 
and~\cite{Davies:2009tsa}
\begin{equation} \label{r1}
   r_1 = 0.3133(23)\,\mathrm{fm}.
\end{equation}

The MILC configurations include vacuum polarization contributions from 
only the lightest three quark flavors, using the ASQTAD discretization. 
Vacuum polarization effects from the heavier $c$~and $b$~quarks are 
easily incorporated into our final results for quark masses and the QCD 
coupling  using perturbation theory.

We computed heavy-quark correlators (\eq{G}) using the HISQ 
discretization~\cite{Follana:2006rc} for a variety of bare heavy-quark 
masses~$am_{0h}$ on the MILC gluon configurations. Our results for the 
reduced moments $R_n$ with $n\!=\!4$--$18$ are given in 
Table~\ref{tab:Rn-MC}.

In Table~\ref{tab:Rn-MC}, we also give masses $am_{\eta_h}$ from the 
simulations for the pseudo-Goldstone meson made from two heavy quarks. 
These were computed using single-exponential fits to $G(t)$ for the 
middle 30\% of~$t$s on the lattice for all configurations except the 
two smallest lattice spacings where we used only 8\% of the~$t$s. We 
have less statistics for the two finest lattice spacings and 
consequently the fits did not work as well for these. We increased the 
statistical errors on our results for $am_{\eta_h}$ by factors of~1.4 
and~2 for the next-to-finest and finest lattice spacings (sets~8 
and~9), respectively, to account for this. The statistical errors here 
are very small and have only a small impact on our final results. We 
also verified our results with multi-exponential fits in every case.

\subsection{Systematic Errors}
As discussed above, our goal is to find values for $\alpha_\msb(\mu)$ 
and $z(\mu/m_h,m_{\eta_h})$ (\eq{z-def}) that make the theoretical 
results from perturbative QCD agree, to within statistical and 
systematic errors, with Monte Carlo simulation ``data'' for the reduced 
moments. We simultaneously analyze results for all of our lattice 
spacings and most of our masses, and for moments with~$4\le n\le10$. We 
focus on these particular moments for our final results since their 
perturbation theory is known to third order.

Systematic errors are larger here than statistical errors, which 
contribute less than~$0.3\%$. We discuss the most important sources of 
systematic error in this section.

\subsubsection{$m_h$ Extrapolations}
We need the  $m_{\eta_h}$ dependence of the mass-ratio function 
$z(\mu/m_h\!=\!3,m_{\eta_h})$ in order to extract $c$~and $b$~masses 
from our simulation (using \eq{m-determination}). We parameterize this 
dependence as follows:
\begin{equation}
   z(\mu/m_h,m_{\eta_h}) = \sum_{j=0}^{N_z} z_j(\mu/m_h) \left(
   \frac{2\Lambda}{m_{\eta_h}}\right)^j,
\end{equation}
where the $z_j$s are determined in our fit.
This is an expansion in the QCD scale, which we take to be
\begin{equation}
   \Lambda=0.5\,\mathrm{GeV},
\end{equation}
divided by~$m_{\eta_h}/2$, which we use as a proxy for the quark mass. 
The expansion is adequate for the range of quark masses used in our 
analysis, where $(2\Lambda/m_{\eta_h})^2$ ranges approximately between 
$1/m_{\eta_b}^2 \!=\! 0.01$ and $(1/m_{\eta_c})^2\!=\!0.1$; the 
singular point $m_h\!=\!0$ is infinitely far away in this 
parameterization. In our fits we keep terms only through order 
$N_z\!=\!4$, but, as we discuss later, our results are unchanged by 
additional terms. On dimensional grounds, we assume \emph{a priori} 
that the coefficients are
\begin{equation} \label{z-prior}
   z_j(3) = 0 \pm 1.
\end{equation}

\subsubsection{Finite-Lattice Spacing Errors}\label{sec:finite-a}
Discretization errors are of order $(am_h)^{2i}\alpha_s$ 
for~$i\!\ge\!1$. We model these by
\begin{equation} \label{Rnlatt-fit}
   R_n^\mathrm{latt} = R_n(\mu,m_{\eta_h},a,N_{am}),
\end{equation}
where: fit function $R_n(\mu,m_{\eta_h},a,N_{am})$ has the double 
expansion
\begin{align} \label{Rnlatt-fcn}
   R_n(\mu,m_{\eta_h},&a,N_{am}) \equiv R_n^\mathrm{cont} / 
\\ \nonumber
   &\left(1+\sum_{i=1}^{N_{am}}
   \sum_{j=0}^{N_{z}} c_{ij}^{(n)}
    \left(\frac{am_{\eta_h}}{2}\right)^{2i}
   \left(\frac{2\Lambda}{m_{\eta_h}}\right)^j
   \right),
\end{align}
the $c_{ij}^{(n)}$s are determined in our fit, $R_n^\mathrm{cont}$ is 
given by~\eq{Rn-eqn},
\begin{equation}
   i+j\le\mathrm{max}(N_{am},N_{z}),
\end{equation}
and again we use~$m_{\eta_h}/2$ in place of the quark mass. This 
expansion allows for finite-$a$ corrections involving 
$(am_{\eta_h}/2)^2$, $(a\Lambda)^2$, and cross terms, with 
$m_{\eta_h}$-dependent coefficients.
We assume \emph{a priori} that
\begin{equation} \label{cijk-prior}
   c_{ij}^{(n)} = 0 \pm 2/n
\end{equation}
which implies smaller $a$~dependence for larger~$n$s. This is expected 
(and obvious in our simulation data) since the reduced moments become 
more infrared as $n$~increases. The exact functional form of the 
$n$~dependence has little effect on our results, as we show later.

In our fits we take $N_z\!=\!4$. While low orders suffice for the 
$2\Lambda/m_{\eta_h}$ expansion, expansion parameter $am_{\eta_h}/2$ 
ranges between~0.3 and~1.1, and higher orders are necessary, especially 
given our tiny statistical errors. We find that our fit results don't  
converge well unless $N_{am}$ is larger than~10--20. Also we have 
difficulty getting good fits if we include data with 
$am_{\eta_h}\!>\!1.95$ from Table~\ref{tab:Rn-MC}. The $am_{\eta_h}/2$ 
expansion may not converge for these last cases and therefore we 
exclude such data from our final analysis.

The fit function has many more fit parameters $c_{ij}^{(n)}$ than we 
have simulation data points when $N_{am}$ is so large. This does not 
cause problems in (Bayesian) constrained fits since the parameters' 
priors (\eq{cijk-prior}) are included in the fit as extra 
data~\cite{Lepage:2001ym}. Each parameter has a prior and therefore we 
always have more data than parameters.

It is, however, very time consuming to fit a function with so many fit 
parameters. Although it is not essential for our analysis, there is a 
trick that greatly accelerates this kind of fit. The idea is to fit a 
modified moment $\bar{R}_n^\mathrm{latt}$ in place of 
$R_n^\mathrm{latt}$ where
\begin{align} \label{Rbar-def}
\bar{R}_n^\mathrm{latt}& \equiv R_n^\mathrm{latt} + \\ \nonumber
&R_n^\mathrm{latt} \sum_{i=\bar{N}_{am}+1}^{N_{am}}\sum_{j=0}^{N_z}
 c_{ij}^{(n)}
 \left(\frac{am_{\eta_h}}{2}\right)^{2i}
\left(\frac{2\Lambda}{m_{\eta_h}}\right)^j.
\end{align}
and $\bar{N}_{am}\!\ll\!N_{am}$. The modified moment is fit with the 
much simpler formula (simpler since $\bar{N}_{am}\!\ll\!N_{am}$)
\begin{equation} \label{Rbar-formula}
   \bar{R}_n^\mathrm{latt} = {R}_n(\mu,m_{\eta_h},a,\bar{N}_{am}).
\end{equation}
where $R_n(\ldots)$ is again given by~\eq{Rnlatt-fcn}. To evaluate 
$\bar{R}_n^\mathrm{latt}$ from \eq{Rbar-def}, we treat the 
coefficients~$c_{ij}^{(n)}$ with~$i\!>\!\bar{N}_{am}$ as new data with 
means and standard deviations specified by the prior,~\eq{cijk-prior}.  
Uncertainties coming from the $c_{ij}^{(n)}$s are combined in 
quadrature with the statistical error in $R_n^\mathrm{latt}$ to obtain 
a new error estimate for~$\bar{R}_n^\mathrm{latt}$ (but leaving the 
central value unchanged). In effect we are increasing the error in the 
reduced moment to account for high-order $(am_{\eta_h}/2)^{2i}$ terms 
omitted from the fit formula~\eq{Rbar-formula}. By choosing 
$\bar{N}_{am} \ll N_{am}$, most of the $am_{\eta_h}/2$ terms are 
incorporated into $\bar{R}_n^\mathrm{latt}$ (\eq{Rbar-def}), where they 
are inexpensive, and relatively few end up in the fit 
function~$\bar{R}_n(\ldots)$ (\eq{Rnlatt-fcn}), where they add 
parameters to the fit and increase its cost. Note that the new errors 
introduce correlations between $\bar{R}_n^\mathrm{latt}$s computed with 
different lattice spacings or quark masses, since the same 
$c_{ij}^{(n)}$s are used for all~$a$s and~$m_{\eta_h}$s. These 
correlations are important and need to be preserved in the fit.

Our procedure, whereby terms are moved out of the fitting function and 
incorporated into new (correlated) errors in the Monte Carlo fit data, 
is generally useful. Somewhat remarkably, final fit results are 
completely (or almost completely) independent of the number of terms 
that are transferred when fits are linear (or almost linear) in the 
associated parameters. (The general theorem from which this result 
follows is proven in the Appendix.) Consequently, in our analysis here, 
we can take $N_{am}$ very large\,---\,say, $N_{am}\!=\!80$\,---\,and 
still have very fast fits by keeping $\bar{N}_{am}$ very small. With 
$N_{am}\!=\!80$ we find, for example, that setting $\bar{N}_{am}\!=\!0$ 
in $\bar{R}_n(\ldots)$ (no terms) gives essentially identical results 
for our quark masses and coupling as setting $\bar{N}_{am}\!=\!30$ 
(140~terms), even though the latter fit requires 22~times more 
computing.  We used this procedure, with $\bar{N}_{am}\!=\!0$, for most 
of our testing and development in this project.

\subsubsection{Truncated Perturbation Theory}

\begin{table}
   \caption{Perturbation theory coefficients ($n_f\!=\!3$) for
   $r_n$~\cite{Chetyrkin:2006xg,
   Boughezal:2006px,Maier:2008he,Maier:2009fz,Kiyo:2009gb}. 
Coefficients
   are defined by $r_n\!=\!1+\sum_{j=1} r_{nj}\almsb^j(\mu)$ for
   $\mu\!=\!m_h(\mu)$. The third-order coefficients are exact for $4\le
   n\le10$. The other coefficients are based upon estimates; we assign
   conservative errors to these.}
   \label{tab:rn-pth}
   \begin{ruledtabular}
   \begin{tabular}{rrrr}
   $n$ & $r_{n1}$ & $r_{n2}$ & $r_{n3}$ \\ \hline
    4 & 0.7427 & $-$0.0577 & 0.0591 \\
    6 & 0.6160 & 0.4767 & $-$0.0527 \\
    8 & 0.3164 & 0.3446 & 0.0634 \\
   10 & 0.1861 & 0.2696 & 0.1238 \\
   12 & 0.1081 & 0.2130 & 0.1(3) \\
   14 & 0.0544 & 0.1674 & 0.1(3) \\
   16 & 0.0146 & 0.1293 & 0.1(3) \\
   18 & $-$0.0165 & 0.0965 & 0.1(3) \\
   \end{tabular}
   \end{ruledtabular}
\end{table}

The perturbative part,
\begin{equation}
   r_n(\alpha_\msb,\mu/m_h) = 1 + \sum_{j=1}^{N_\mathrm{pth}}
   r_{nj}(\mu/m_h) \alpha_\msb^j(\mu),
\end{equation}
of the reduced moments is known at best through third order. We present 
coefficients~$r_{nj}$ through~$j\!=\!3$ in 
Table~\ref{tab:rn-pth}~\cite{Chetyrkin:2006xg,Boughezal:2006px,Maier:2008he,Maier:2009fz,Kiyo:2009gb}; the values for $n\!=\!4$--10 are exact, while $r_{n3}$ is estimated for the others. In our fits we include higher-order terms by treating the coefficients of these terms as fit parameters with prior
\begin{equation} \label{rn-prior}
   r_{nj}(1) = 0 \pm 0.5
\end{equation}
for any coefficient that hasn't been computed in perturbation theory. 
We set $N_\mathrm{pth}\!=\!6$ since then contributions from still 
higher orders should be less than~$0.1$\% (and setting 
$N_\mathrm{pth}\!=\!8$ doesn't change our results).

The perturbative coefficients for $\mu/m_h\!=\!1$ 
(Table~\ref{tab:rn-pth}) are small and relatively uncorrelated from 
order-to-order. This is less true for $\mu/m_h\!=\!3$, which is where 
we wish to work (see Section~II), because of $\log(\mu/m_h)^m$ terms. 
In order to capture these effects, we use renormalization group 
equations to express the $r_{nj}(3)$~coefficients (for all $j\le 
N_\mathrm{pth}$) in terms of the $r_{nj}(1)$ coefficients and 
$\log(\mu/m_h)$, and substitute the results from Table~\ref{tab:rn-pth} 
for $j\le3$ and from the prior (\eq{rn-prior}) for~$j>3$. This 
procedure generates (correlated) priors for the unknown coefficients at 
$\mu/m_h\!=\!3$ that properly account for renormalization-group 
logarithms.

\subsubsection{$\alpha_\msb$ Evolution}
As discussed above, we fix the ratio of $\mu/m_h(\mu)$ in our analysis. 
This means that the renormalization scale $\mu$ varies over a wide 
range of values for the different $m_h$s we use. The coupling constant 
$\alpha_\msb(\mu)$ used in the perturbative expansions for the $r_n$s 
is specified at $\mu\!=\!5$\,GeV by fit parameter $\alpha_0$, with 
prior
\begin{equation} \label{alpha0-prior}
   \alpha_0 = 0.20 \pm 0.01,
\end{equation}
where
\begin{equation} \label{alpha0-def}
   \alpha_0 \equiv \alpha_\msb(5\,\mathrm{GeV},n_f=3).
\end{equation}
The prior corresponds to $\alpha_\msb(M_Z)\!=\!0.118(3)$\,---\,a very 
broad range, which means that the prior has little impact on our final 
fit results. The coupling value at any scale $\mu\ne5$\,GeV is obtained 
by integrating (numerically) the QCD evolution equation for 
$\alpha_\msb(\mu)$ starting with value~$\alpha_0$ at scale~$5$\,GeV. We 
use the $\msb$~beta function through sixth order in~$\alpha_\msb$,
\begin{align} \label{alpha-evol}
   \mu^2 \frac{d\alpha_\msb(\mu)}{d\mu^2} =&
   -\beta_0 \alpha_\msb^2 - \beta_1 \alpha_\msb^3
   -\beta_2 \alpha_\msb^4 \\ \nonumber
   & - \beta_3 \alpha_\msb^5
   -\beta_4 \alpha_\msb^6,
\end{align}
where $\beta_0\ldots\beta_3$ are known from perturbation theory and 
$\beta_4$ is taken as a fit parameter with prior
\begin{equation} \label{beta-prior}
   \beta_4 = 0 \pm \sigma_\beta
\end{equation}
where $\sigma_\beta$ is the root-mean-square average 
of~$\beta_0\ldots\beta_3$~\cite{van Ritbergen:1997va,Czakon:2004bu}. We 
include this last term to estimate the uncertainties in our final 
results caused by unknown terms in the beta function.

Our simulations include vacuum polarization effects from only the three 
lightest quarks. We use perturbation theory, together with the $c$~and 
$b$~masses that come out of our analysis, to incorporate vacuum 
polarization effects from the heavier quarks into our final results for 
the masses and QCD coupling (using formulas 
from~\cite{Chetyrkin:1997dh,Vermaseren:1997fq} to add the $c$~and 
$b$~quarks at scales $\mu\!=\!m_c$ and~$m_b$, respectively).

\subsubsection{Nonperturbative Condensates}
As discussed in our previous paper, nonperturbative effects dominate 
the reduced moments when $n$~is large. The dominant nonperturbative 
contribution, which is from the gluon condensate, is quite small, 
however, for the range of $n$s and quark masses we use here. We correct 
for it by
replacing
\begin{equation} \label{cond-corrn}
   {R}_n^\mathrm{latt} \to
   {R}_n^\mathrm{latt} \left( 1 +
   d_n \frac{\langle \alpha_s G^2/\pi\rangle}{(2m_h)^4}
   \right)
\end{equation}
where $d_n$ is computed to leading order in perturbation 
theory~\cite{Broadhurst:1994qj} with $m_h\!=\!m_h(m_h)$, which we 
approximate by~$m_{\eta_h}/2.27$. We take
\begin{equation}
   \langle \alpha_s G^2/\pi\rangle = 0\pm0.012\,\mathrm{GeV}^4,
\end{equation}
which covers the range of most current estimates~\cite{cond-footnote}. 
The correction factor in \eq{cond-corrn} adds (slightly) to the error 
in~${R}_n^\mathrm{latt}$
(and introduces new correlations between different moments, since the 
same $\langle \alpha_s G^2/\pi\rangle$ is assumed for every moment, 
lattice spacing and quark mass).

\subsubsection{Finite Volume Errors}
We expect small errors due to the fact that our simulation lattices are 
only about~2.5\,fm across. We allow for the possibility of 
finite-volume errors by replacing
\begin{equation} \label{fvol-corrn}
   {R}_n^\mathrm{latt} \to
   {R}_n^\mathrm{latt} \left( 1 +
   f_n \frac{\Delta {R}_n^\mathrm{pth}}{{R}_n^\mathrm{pth}}
   \right)
\end{equation}
where $\Delta {R}_n^\mathrm{pth}$ is the finite volume error in 
leading-order perturbation theory and
\begin{equation}
   f_n = 0 \pm 0.5.
\end{equation}

The true finite-volume errors are expected to be smaller, because of 
quark confinement, than the perturbative errors that we use to model 
them here. We verified this by running two sets of simulations that 
were identical except for the spatial volume (gluon configuration 
sets~4 and~5 in Table~\ref{tab:cfg}). The differences between the two 
simulations are smaller than our statistical errors, but the 
statistical errors are much smaller than our estimate above. Our error 
estimate here is very conservative, but has negligible impact on our 
final results.

\subsubsection{Sea-Quark Masses}
The sea-quark masses used in our simulations are not exactly correct. 
To correct for this we replace
\begin{equation} \label{sea-corrn}
   {R}_n^\mathrm{latt} \to
   {R}_n^\mathrm{latt} \left( 1 +
   g_n \frac{2\delta m_l + \delta m_s}{m_s}
   \right)
\end{equation}
where $\delta m_l$ and $\delta m_s$ are the errors in the $u/d$ and 
$s$~masses (see \cite{Davies:2009tsa} for more details), respectively, 
and
\begin{equation}
   g_n = 0 \pm 0.01.
\end{equation}
This correction introduces (correlated) errors into 
the~${R}_n^\mathrm{latt}$s that are of order 0.5--1\%. Direct 
comparison of results from configuration sets~6 and~7 (or~1--2 
and~3--4) in Table~\ref{tab:cfg} suggests that sea-quark mass effects 
are no larger than 0.1\%, so our error estimate is conservative.

We have only included the leading dependence on the sea-quark mass, 
which comes from nonperturbative (chiral) effects. Quadratic terms from 
perturbation theory and other nonperturbative sources are negligible.

\section{Analysis and Results}
We have computed reduced moments for 30~different sets of lattice 
spacing, lattice volume and quark masses (Table~\ref{tab:Rn-MC}). To 
extract quark masses and the QCD coupling, we fit moments with $4\le 
n\le10$ from 22~of these parameter sets (the ones with 
$am_{\eta_h}\le1.95$)\,---\,88~pieces of simulation data in all. In 
this section we first describe the fitting method used to extract the 
masses and coupling, and then we review our results.

\subsection{Constrained Fits}
We analyze all four $R_n$s for all 22~parameter sets simultaneously 
using a constrained fitting procedure based upon Bayesian 
ideas~\cite{Lepage:2001ym}. In this procedure we minimize an augmented 
$\chi^2$ function of the form
\begin{equation}
   \chi^2 = \sum_{in,jm} \Delta R_{ni}\,\,
   (\sigma^{-2}_R)_{in,jm}\,\, \Delta R_{mj} + \sum_\xi 
\delta\chi^2_\xi
\end{equation}
where:
\begin{equation}
   \Delta {R}_{ni} \equiv
   {R}_{ni}^\mathrm{latt}-{R}_n(\mu_i,m_{\eta_hi},a_i,N_{am});
\end{equation}
the $R_{n}^\mathrm{latt}$ come from Table~\ref{tab:Rn-MC} with 
corrections from Eqs.~(\ref{cond-corrn}), (\ref{fvol-corrn}) 
and~(\ref{sea-corrn}); fit function ${R}_n(\ldots)$ is defined 
by~\eq{Rnlatt-fcn}; and $\sigma^2_R$ is the error covariance matrix for 
the ${R}_{n}^\mathrm{latt}$. The sums $i,j$ are over the 22~sets of 
lattice spacings and quark masses; the sums $n,m$ range over of the 
moments~$4,6,8,10$.

Function ${R}_n(\mu_i,m_{\eta_hi},a_i,N_{am})$ depends upon a large 
number of parameters, all of which are varied in the fit to 
minimize~$\chi^2$. Priors~$\delta \chi^2_\xi$ are included for each of 
these:
\begin{itemize}
   \item parameters $z_j$, with  prior~\eq{z-prior}, from the 
$1/m_{\eta_h}$ expansion of $z(\mu/m_h,m_{\eta_h})$;
   \item parameters $c_{ij}^{(n)}$, with prior \eq{cijk-prior}, from 
the finite-lattice spacing corrections;
   \item unknown perturbative coefficients $r_{nj}$, with prior 
\eq{rn-prior} (evolved to $\mu/m_h\!=\!3$);
   \item coupling parameter $\log(\alpha_0)$, with 
prior~\eq{alpha0-prior};
   \item $\beta_4$ in the QCD $\beta$-function, with 
prior~\eq{beta-prior};
   \item lattice spacings $a_i$ for each gluon configuration set, with 
priors specified by simulation results for $r_1/a$ 
(Table~\ref{tab:cfg}) and the current value for~$r_1$ (\eq{r1});
   \item values for $am_{\eta_hi}$, with priors specified by our 
simulation results (Table~\ref{tab:Rn-MC}). \end{itemize}
The renormalization scales $\mu_i$ are obtained from the ratio 
$\mu/m_h\!=\!3$, simulation results for $m_{\eta_h}$, and~\eq{z-def}. 
We take $N_{am}\!=\!30$ for our final results.

\subsection{Results}\label{sec:results}
\begin{figure}
   \includegraphics[scale=1.0]{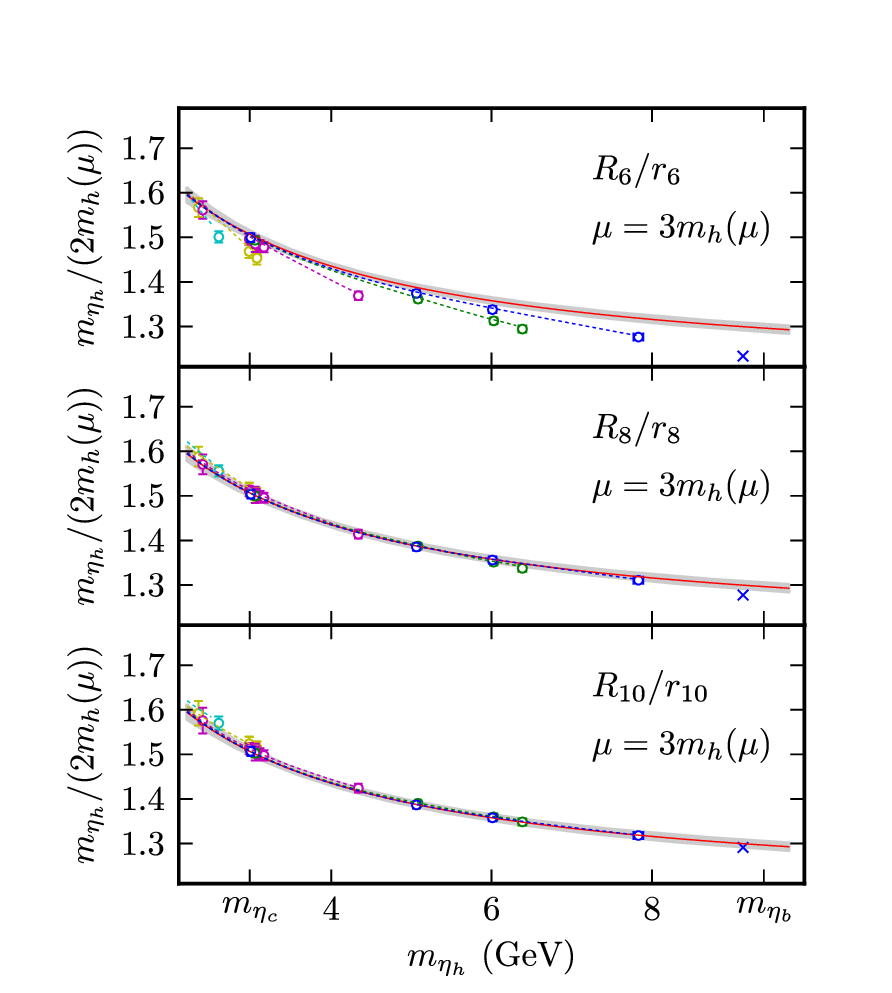}
   \caption{Function $z(\mu/m_h\!=\!3,m_{\eta_h})\!\equiv\! 
m_{\eta_h}/(2m_h)$ as a function of $m_{\eta_h}$. The solid line, plus 
gray error envelope, shows the $a\!=\!0$ extrapolation obtained from 
our fit. This is compared with simulation results for $R_n/r_n$ for 
$n\!=\!6,8,10$ from our 5~different lattice spacings, together with the 
best fits (dashed lines) corresponding to those lattice spacings. 
Dashed lines for smaller lattice spacings extend further to the right. 
The points marked by an ``\textsf{x}'' are for the largest mass we 
tried (last line in Table~\ref{tab:Rn-MC}); these are not included in 
the fit because $am_{\eta_h}$ is too large.
Finite-$a$ errors become very small for the larger-$n$ moments, causing 
points from different lattice spacings to overlap.}
   \label{fig:z}
\end{figure}

We fit our simulation data for the reduced moments~$R_n^\mathrm{latt}$ 
(Table~\ref{tab:Rn-MC}) using fit function $R_n(\ldots)$ 
(\eq{Rnlatt-fcn}) with~$N_{am}\!=\!30$, as discussed in the previous 
section. The best-fit values for parameters~$z_j$ give us the 
mass-ratio function~$z(\mu/m_h\!=\!3,m_{\eta_h})$ (\eq{z-def}), which 
we plot in Figure~\ref{fig:z}. We also show our simulation results 
there for ${R}_n^\mathrm{latt}/r_n$, together with the best-fit lines 
for each lattice spacing. Results are shown for the three moments that 
depend upon~$z$, 5~different lattice spacings, and quark masses ranging 
from below the $c$~mass almost to the $b$~mass. The simulation data 
were all fit simultaneously, using the same functions~$z(3,m_{\eta_h})$ 
and $\alpha_\msb(\mu)$ (with $\mu\!=\!3m_{\eta_h}/(2z)$) for all 
moments. The fits are excellent, with $\chi^2/88=0.19$ for the 
88~pieces of simulation data we fit.

Evaluated at $m_{\eta_c}\!=\!2.985(3)$\,GeV~\cite{etac-ref}, the 
mass-ratio function is $z(3,m_{\eta_c})\!=\!\zmc$. Combining this 
with~\eq{m-determination} and perturbation theory, we can obtain the 
following results for the $\msb$ $c$-quark mass at different scales:
\begin{align}
   m_c(3m_c,n_f=3) &= \mcmu\,\mathrm{GeV}, \\ \nonumber
   m_c(3\,\mathrm{GeV},n_f=4) &= \mc\,\mathrm{GeV}, \\ \nonumber
   m_c(m_c,n_f=4) &= \mcmc\,\mathrm{GeV}.
\end{align}

Similarly at $m_{\eta_b}\!=\!9.395(5)$\,GeV~\cite{etab-ref}, the 
mass-ratio function is $z(3,m_{\eta_b})\!=\!\zmb$, and we obtain the 
following results for the $\msb$ $b$-quark mass at different scales:
\begin{align}
   m_b(3m_b,n_f=3) &= \mbmu\,\mathrm{GeV}. \\ \nonumber
   m_b(10\,\mathrm{GeV},n_f=5) &= \mb\,\mathrm{GeV}, \\ \nonumber
   m_b(m_b,n_f=5) &= \mbmb\,\mathrm{GeV}.
\end{align}

Note that the ratio $m_b(\mu,n_f)/m_c(\mu,n_f)$ is independent of $\mu$ 
and $n_f$. We obtain the following result for this mass ratio:
\begin{equation} \label{mbmc}
   m_b/m_c = \mbmc
\end{equation}

\begin{figure}
   \includegraphics[scale=1.0]{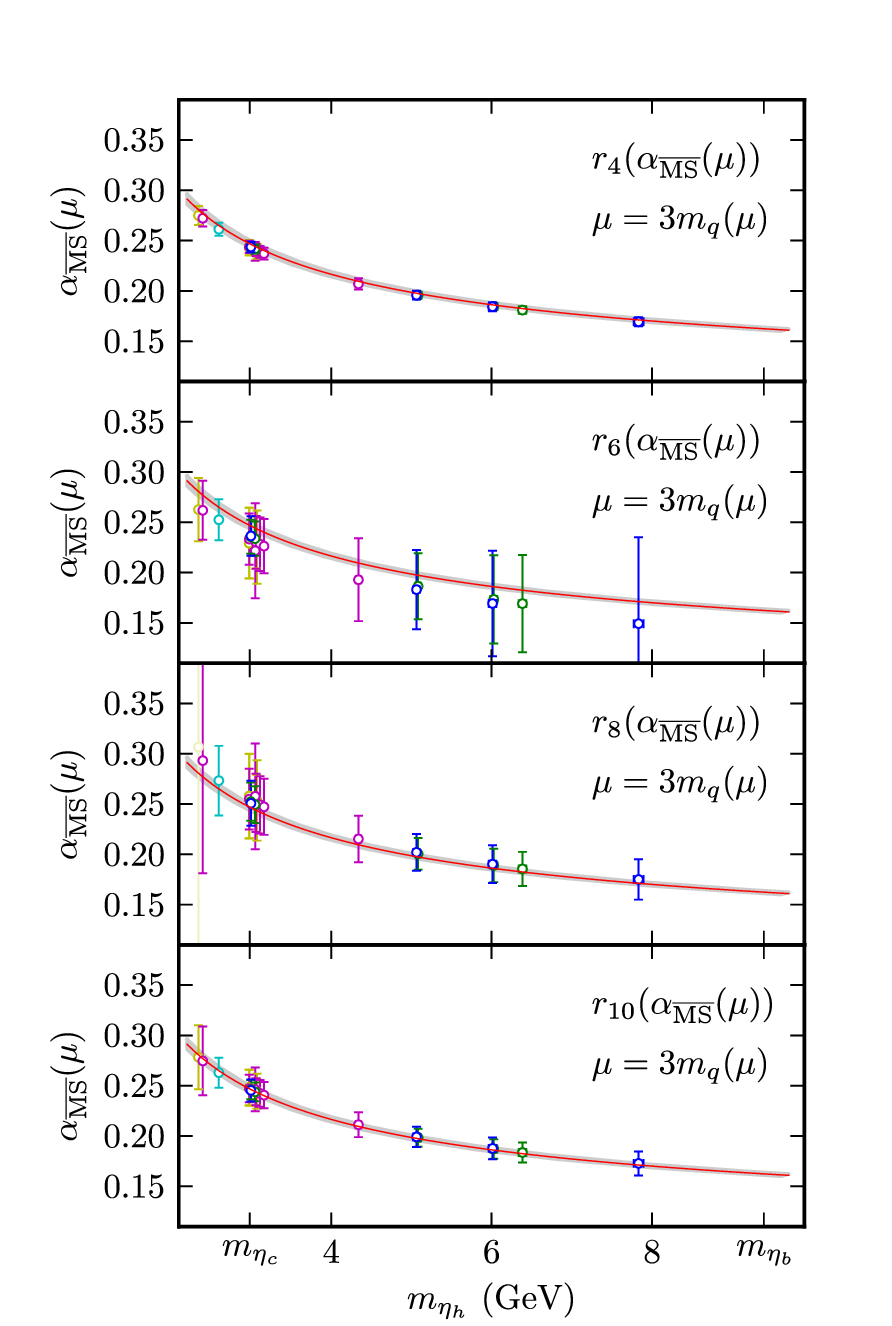}
   \caption{QCD coupling $\alpha_\msb(\mu,n_f\!=\!3)$ as a function of 
$m_{\eta_h}$ where $\mu\!=\!3m_h$. The solid line, plus gray error 
envelope, shows the best-fit coupling from our fit when perturbative 
evolution is assumed. The data points are values of $\alpha_\msb$ 
extracted from individual simulation results for $R_n$ after 
extrapolating to $a\!=\!0$ and dividing out~$z(3,m_{\eta_h})$ 
($n\!>\!4$). Results are given for moments $n\!=\!4$--10 and all 
5~lattice spacings. Several points from different lattice spacings 
overlap in these plots.}
   \label{fig:alphamsb}
\end{figure}

The other important output from our fit is a value for the parameter
\begin{equation}
   \alpha_0 \equiv \alpha_\msb(5\,\mathrm{GeV},n_f=3) = \alphafit.
\end{equation}
To compare with other determinations of the coupling, we add vacuum 
polarization corrections from the $c$~and $b$~quarks, using the masses 
above, and evolve to the $Z$-meson mass~\cite{van Ritbergen:1997va, 
Czakon:2004bu,Chetyrkin:1997dh,Vermaseren:1997fq}:
\begin{equation} \label{almz}
   \alpha_\msb(M_Z,n_f=5) = \alphaMZ.
\end{equation}
Figure~\ref{fig:alphamsb} shows how consistent our simulation results 
are with the theoretical curve for $\alpha_\msb(\mu,n_f\!=\!3)$ 
corresponding to our value for~$\alpha_0$. For this figure we extracted 
values for $\alpha_\msb$ from each~$R_n$ separately by dividing out the 
$a^2$~dependence and $z(3,m_{\eta_h})$ using our best-fit parameters, 
and then solving for $\alpha_\msb$ by matching with perturbation theory 
for~$r_n$. (In our fit, of course, we fit all $R_n$s simultaneously to 
obtain a single~$\alpha_\msb$ for all of them.)

The dominant sources of error for our results are listed in 
Table~\ref{tab:error-budget}. The largest uncertainties come from: 
extrapolations to $a\!=\!0$, especially for quantities involving 
$b$~quarks; unknown higher-order terms in perturbation theory, 
especially for quantities involving $c$~quarks; statistical 
fluctuations; extrapolations  in the heavy quark mass, especially for 
quantities involving $b$~quarks; and uncertainties in static-quark 
parameters $r_1/a$ and $r_1$.The pattern of errors is as expected in 
each case. The nonperturbative contribution from the gluon condensate 
is negligible except for $m_c$, again as expected; and errors due to 
mistuned sea-quark masses, finite volume errors, and uncertainties in 
$\msb$ coupling and mass evolution are negligible ($<\!0.05\%$).
\begin{table}
   \caption{Sources of uncertainty for the QCD coupling and mass 
determinations in this paper. In each case the uncertainty is given as 
a percentage of the final value.}
   \label{tab:error-budget}
\begin{ruledtabular}\begin{tabular}{ccccc}
 & $\alpha_\msb(M_Z)$ & $m_b(10)$ & $m_b/m_c$
 & $m_c(3)$ \\ \hline
           $a^2$ extrapolation &  0.2\% &  0.6\% &  0.5\% &  0.2\%\\
           perturbation theory &  0.5 &  0.1 &  0.5 &  0.4\\
           statistical errors  &  0.1 &  0.3 &  0.3 &  0.2\\
           $m_h$ extrapolation &  0.1 &  0.1 &  0.2 &  0.0\\
             errors in $r_1$ &  0.2 &  0.1 &  0.1 &  0.1\\
           errors in $r_1/a$ &  0.1 &  0.3 &  0.2 &  0.1\\
           errors in $m_{\eta_c}$,$m_{\eta_b}$
                           &  0.2 &  0.1 &  0.2 &  0.0\\
        $\alpha_0$ prior &  0.1 &  0.1 &  0.1 &  0.1\\
              gluon condensate &  0.0 &  0.0 &  0.0 &  0.2\\ \hline
               Total  &  0.6\% &  0.7\% &  0.8\% &  0.6\%
\end{tabular}\end{ruledtabular}
\end{table}

\begin{figure}
   \includegraphics[scale=1.0]{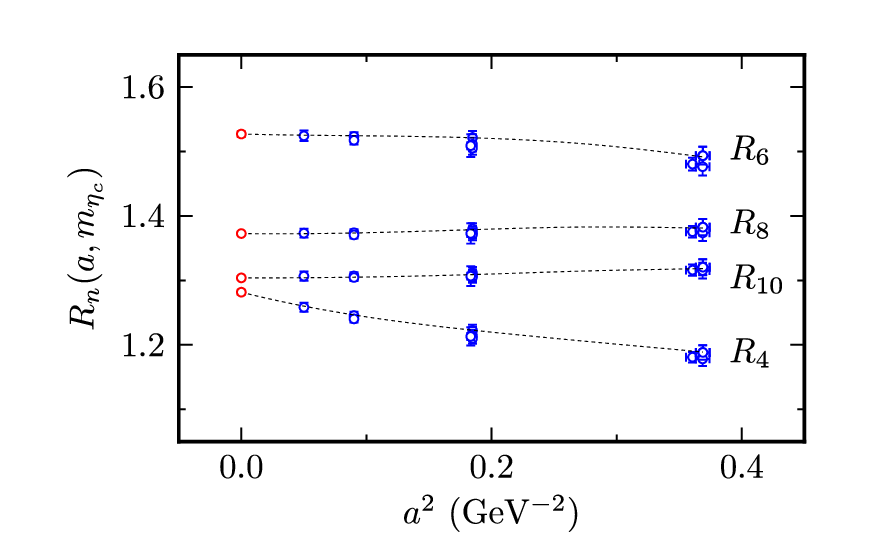}
   \caption{Lattice spacing dependence of ${R}_n$ for masses 
$m_{\eta_h}$ within~5\% of~$m_{\eta_c}$ and moments 
$n\!=\mathbf{4},6,8,10$. The dashed lines show our fit for the average 
of these masses, and the points at $a\!=\!0$ are the continuum 
extrapolations of our data.}
   \label{fig:rn-a2}
\end{figure}

The $a^2$~extrapolations of our data are not large. This is illustrated 
for $m_h\approx m_c$ in Figure~\ref{fig:rn-a2}, which shows the 
$a^2$~dependence of the reduced moments. The smallest two lattice 
spacings are sufficiently close to $a\!=\!0$ that the extrapolation is 
almost linear from those points. The $a\!=\!0$ extrapolated values we 
obtain here for the $R_n$ agree to within (smaller) errors with those 
in our previous paper: here we get 1.282(4), 1.527(4), 1.373(3), 
1.304(2) with $n\!=\!4,6,8,10$, respectively, for the masses used in 
the figure.

We tested the stability of our analysis in several ways:
\begin{itemize}
   \item \emph{Vary perturbation theory:} We chose $\mu\!=\!3m_h$ in 
order to keep scales large and $\alpha_\msb(\mu)$ small. Our results 
are quite insensitive to~$\mu$, however. Choosing $\mu\!=\!m_h$, for 
example, shifts none of our results by more than $0.2\sigma$, and 
leaves all errors unchanged except for $m_c(3)$, where the error 
increases by a third. Taking $\mu\!=\!9m_h$ shifts results by less than 
$0.4\sigma$, and reduces the $m_c$ error by a third, leaving others 
only slightly reduced. Adding more terms to the perturbative expansions 
($N_\mathrm{pth}\!=\!6\to8$) also has essentially no effect on the 
results. The prior for the unknown perturbative coefficients 
(\eq{rn-prior}) is twice as wide as suggested by our simulation results 
(using the empirical Bayes criterion~\cite{Lepage:2001ym}); we choose 
the larger width to be conservative.

   \item \emph{Include more/fewer finite-$a$ corrections:} We set 
$N_{am}\!=\!30$ for our results above. Using $N_{am}\!=\!15$ gives 
results that differ by less than $0.5\sigma$ for $m_b$ and much less 
for the other quantities. Much larger $N_{am}$s can be tested easily 
using the trick described in Section~\ref{sec:finite-a}. For example, 
replacing $R_n^\mathrm{latt}$ by $\bar{R}_n^\mathrm{latt}$ 
(\eq{Rbar-def}) with $N_{am}\!=\!80$ and $\bar{N}_{am}\!=\!30$ gives 
results that are essentially identical to those above. As discussed 
above, taking $\bar{N}_{am}\!=\!0$ with the same~$N_{am}$ also gives 
the same results and is 22~times faster (see the appendix for further 
discussion).

   \item \emph{Change $n$ dependence of finite-$a$ corrections:} 
Replacing the $n$-dependent prior for the expansion coefficients 
(\eq{cijk-prior}) by the $n$-independent prior~$0\pm 0.5$ causes 
changes that are less than~$0.3\sigma$. The width of the original prior 
is optimal according to the empirical Bayes criterion\,---\,that is, it 
is the width suggested by the size of finite-$a$ deviations observed in 
our simulation data.

   \item \emph{Add more/fewer $\Lambda/m{\eta_h}$ terms in $z$:} 
Increasing the number of terms in the expansion for $z$ 
from~$N_z\!=\!4$ to~6 changes nothing by more than $0.1\sigma$. 
Decreasing to~$N_z\!=\!3$ also has no effect. Again the width of the 
prior is optimal according to the empirical Bayes criterion.

   \item \emph{Include more/fewer moments:} Keeping all moments $4\le 
n\le18$ changes nothing by more than $0.5\sigma$ and reduces errors 
slightly for everything other than~$m_b$, where the errors are cut 
almost in half: $m_b(10)\!=\!3.623(15)$\,GeV or 
$m_b(m_b)\!=\!4.170(13)$\,GeV, both for~$n_f\!=\!5$. We continue to 
restrict ourselves to moments with $n\!\le\!10$ because these are the 
only moments for which we have exact third-order perturbation theory. 
Keeping just $n\!=\!4,6$ gives almost identical results for $m_c$ and 
$\alpha_\msb$, with almost the same errors, but doubles the error on 
$m_b$.

   \item \emph{Omit simulation data:} The coarsest two lattice spacings 
(configuration sets 1--5) affect our results only weakly. Leaving these 
out shifts no result by more than $0.5\sigma$ and leaves errors almost 
unchanged. Leaving out the smallest lattice spacing, however, increases 
errors significantly (almost double for $\alpha_\msb$), while still 
shifting central values by less than $0.5\sigma$.

   \item \emph{Add large masses:} Including cases with 
$am_{\eta_h}>1.95$ from Table~\ref{tab:Rn-MC} leads to poor fits. The 
excluded data, however, do not deviate far from the best-fit lines. For 
example, the points marked with an~``\textsf{x}'' in Figure~\ref{fig:z} 
are for the largest mass we studied, corresponding to 
$m_{\eta_h}\!=\!9.15$\,GeV (last line in Table~\ref{tab:Rn-MC}). 
Although $am_{\eta_h}$ is too large for this case to be included in our 
fit, the values of $R_n/r_n$ are only slightly below the fit results.
\end{itemize}

\section{Nonperturbative $m_b/m_c$}
It is possible to extract the ratio of quark masses $m_b/m_c$ directly, 
without using the moments and without using perturbation theory. This 
provides an excellent nonperturbative check on our results from the 
moments.

Ratios of quark masses are UV-cutoff independent and therefore the 
ratio of $\msb$ masses
\begin{equation}
   \frac{m_b(\mu,n_f)}{m_c(\mu,n_f)} = \frac{m_{0b}}{m_{0c}} + 
\order(\alpha_s a^2m_b^2)
\end{equation}
for any~$\mu$ and~$n_f$, where $m_{0b}$ and $m_{0c}$ are the bare quark 
masses in the lattice quark action that give correct masses for 
the~$\eta_c$ and~$\eta_b$, respectively. We obtain accurate mass ratios 
from this relationship by extrapolating to $a\!=\!0$. We used such a 
method recently to determine $m_c/m_s$~\cite{Davies:2009ih}.

Here we have to modify our earlier method slightly because we
cannot reach the $b$-quark mass directly, but rather must
simultaneously extrapolate to the $b$~mass
and the continuum limit. This is most simply done by determining the 
functional dependence of the ratio
\begin{equation}
   w(m_{\eta_h},a) \equiv \frac{2m_{0h}}{m_{\eta_h}}
\end{equation}
on the $\eta_h$~mass and the lattice spacing. The ratio of 
$\msb$~masses is then given by the experimental masses of the $\eta_c$ 
and $\eta_b$ and the equation:
\begin{equation}\label{eq:mcmb-nonpert}
   \frac{m_b(\mu,n_f)}{m_c(\mu,n_f)} =
\frac{m_{\eta_b}^\mathrm{exp}\,w(m_{\eta_b}^\mathrm{exp},0)}{m_{\eta_c}^\mathrm{exp}\,w(m_{\eta_c}^\mathrm{exp},0)}.
\end{equation}
It might seem simpler to fit~$m_{0h}$ directly, rather than the 
ratio~$w$; but using~$w$ significantly reduces the $m_{\eta_h}$ 
dependence (and therefore our extrapolation errors), and also makes our 
results quite insensitive to uncertainties in our values for the 
lattice spacing.

We parameterize function~$w$ with an expansion modeled after the one we 
used to fit the moments:
\begin{align}
   w(m_{\eta_h},&a) = Z_m(a)\,
   \left(1 + \sum_{n=1}^{N_w} w_n
    \left(\frac{2\Lambda}{m_{\eta_h}}\right)^n\right)
   /
   \\ \nonumber
   &\left(1 + \sum_{i=1}^{N_{am}}
   \sum_{j=0}^{N_{w}} c_{ij} \left(\frac{am_{\eta_h}}{2}\right)^{2i}
   \left(\frac{2\Lambda}{m_{\eta_h}}\right)^j
   \right),
\end{align}
where, as for the moments,
\begin{equation}
   i+j\le\mathrm{max}(N_{am},N_{w}).
\end{equation}
Coefficients $c_{ij}$ and $w_n$ are determined by fitting function 
$w(m_{\eta_h},a)$ to the values of $2am_{0h}/(am_{\eta_h})$ from 
Table~\ref{tab:Rn-MC}. The fit also determines the parameters~$Z_m(a)$, 
one for each lattice spacing, which account for the running of the bare 
quark masses between different lattice spacings.

The finite-$a$ dependence is smaller here than for the moments, because 
the $\eta_h$ is nonrelativistic (finite-$a$ errors are suppressed by 
additional powers of $v/c$~\cite{Follana:2006rc}), and the variation 
with $m_{\eta_h}$ stronger (twice that of $z(3,m_{\eta_h})$). So here 
we use priors
\begin{align}
   c_{ij} &= 0\pm0.05 \\ \nonumber
   w_n &= 0\pm4 \\ \nonumber
   Z_m(a) &= 1\pm0.5
\end{align}
with $N_w\!=\!8$. We again take $N_{am}\!=\!30$, although identical 
results are obtained with $N_{am}=15$.

\begin{figure}
   \includegraphics[scale=1.0]{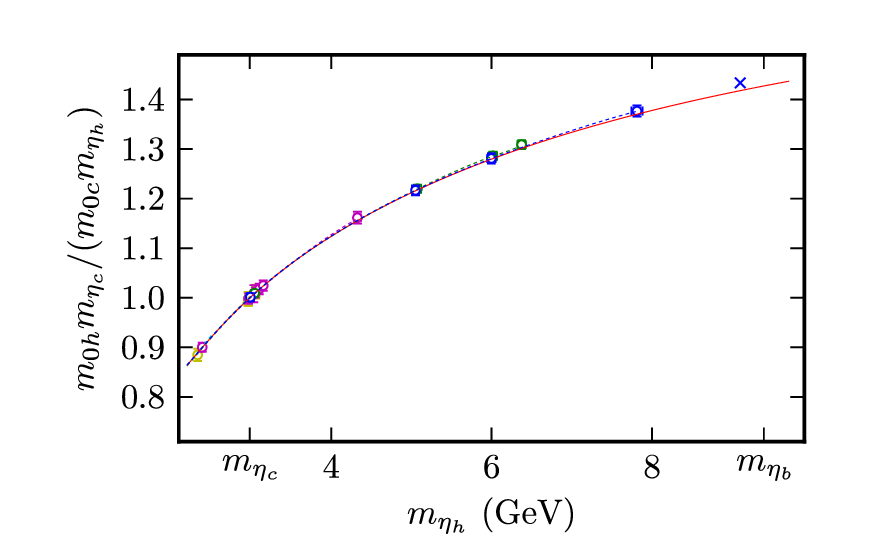}
   \caption{Ratio $m_{0h}/m_{\eta_h}$ divided by $m_{0c}/m_{\eta_c}$ 
(which we approximate by $w(m_{\eta_c},a)/2$ from our fit) as a 
function of $m_{\eta_h}$. The solid line shows the 
$a\!=\!0$~extrapolation obtained from our fit. This is compared with 
simulation results for our 4~smallest lattice spacings, together with 
the best fits (dashed lines) corresponding to those lattice spacings. 
The point marked by an ``\textsf{x}'' is for the largest mass we tried 
(last line in Table~\ref{tab:Rn-MC}); this was not included in the fit 
because $am_{\eta_h}$ is too large.}
   \label{fig:m0-m}
\end{figure}

Our fit results are illustrated by Figure~\ref{fig:m0-m} which plots 
the ratio $m_{0h}/m_{\eta_h}$ divided by $m_{0c}/m_{\eta_c}$ for a 
range of $\eta_h$~masses. Our data for different lattice spacings is 
compared with our fit, and with the $a\!=\!0$~limit of our fit (solid 
line). The fit is excellent, with $\chi^2/22\!=\!0.42$ for the 
22~pieces of data we fit (we again exclude cases with 
$am_{\eta_h}\!>\!1.95$). Using the $\eta_c$ and $\eta_b$ masses from 
Section~\ref{sec:results}, and \eq{eq:mcmb-nonpert} with the best-fit 
values for the parameters, we obtain finally
\begin{align} \label{mbmc-nonpert}
   \frac{m_{0b}}{m_{0c}} &\to 4.49(4) \quad\quad\mbox{as $a\!\to\!0$}
   \\ \nonumber
   &= \frac{m_b(\mu,n_f)}{m_c(\mu,n_f)},
\end{align}
which agrees well with our result from the moments (\eq{mbmc}).

\section{$\alpha_\msb$ from Wilson Loops}
\begin{figure}
   \includegraphics[scale=1.0]{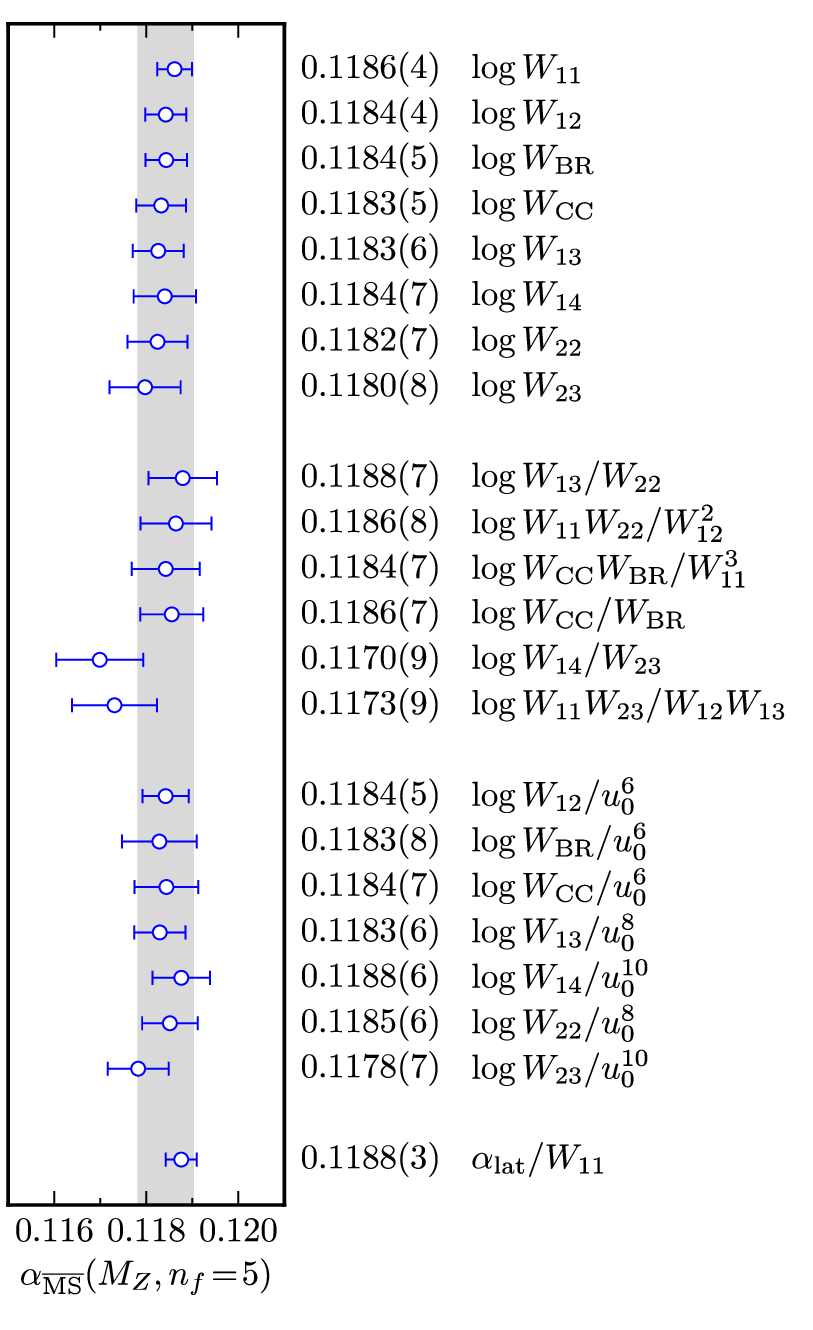}
   \caption{Updated values for the 5-flavor $\alpha_\msb$ at the 
$Z$-meson mass from each of 22~different short-distance quantities 
built from Wilson loops. The gray band indicates a composite average, 
0.1184(6). $\chi^2$ per data point is~0.3.}
   \label{wilson-alpha}
\end{figure}

In a recent paper~\cite{Davies:2008sw}, we presented a very accurate 
determination of the QCD coupling from simulation results for Wilson 
loops. Here we want to compare those results to the value we obtain 
from heavy-quark correlators. First, however, we must update our 
earlier analysis to take account of the new value 
for~$r_1$~\cite{Davies:2009tsa} given in~\eq{r1} and improved values 
for~$r_1/a$~\cite{MILC} given in Table~\ref{tab:cfg}. (The Wilson-loop 
paper uses some additional configuration sets: from Table~II in that 
paper, sets 1, 6, 9, and 11 whose new $r_1/a$s are 1.813(8), 2.644(3), 
5.281(8) and 5.283(8), respectively.) We have rerun our earlier 
analysis, updating $r_1$, $r_1/a$, and the $c$ and $b$~masses. The 
results are shown in Figure~\ref{wilson-alpha}. Combining results as in 
the earlier paper we obtain a final value from the Wilson-loop 
quantities of
\begin{equation}
   \alpha_\msb(M_Z,n_f=5) = 0.1184(6),
\end{equation}
with $\chi^2/22 \!=\! 0.3$ for the 22~quantities in the figure. This 
agrees very well with the result in the earlier paper, 
$\alpha_\msb(M_Z)\!=\!0.1183(8)$, but has a slightly smaller error, as 
expected given the smaller error in~$r_1$. This new value also agrees 
well with our very different determination from heavy-quark 
correlators~(\eq{almz}). A breakdown of the error into its different 
sources can be found in Table~IV of~\cite{Davies:2008sw} (reduce the 
$r_1$ and $r_1/a$ errors in that table by half to account for the 
improved values used here).

\section{Conclusions}
In this paper, we  improve significantly on our previous determinations 
of the QCD coupling and $c$-quark mass from heavy-quark correlators. 
This is principally due to the inclusion of a new, smaller lattice 
spacing in our analysis. We also generated results for a variety of 
quark masses near~$m_c$, allowing us to interpolate more accurately to 
the physical value of~$m_c$. New third-order perturbation theory 
makes~$R_{10}$ as useful now as $R_4$, $R_6$, and $R_8$ were in the 
earlier paper. Finally, in this paper, we fit multiple moments 
simultaneously, determining consistent values simultaneously for both 
the QCD coupling and the quark masses for all moments. Previously we 
examined each moment or ratio of moments independently, extracting 
$m_c$s or $\alpha_\msb$s independently of each other. Our new results,
\begin{align}
   m_c(3\,\mathrm{GeV},n_f=4) &= \mc\,\mathrm{GeV} \\ \nonumber
   \alpha_\msb(M_Z,n_f=5) &= \alphaMZ,
\end{align}
agree well with our older results of 0.986(10)\,GeV and 0.1174(12), 
respectively~\cite{mcjj}.

\begin{figure}
   \includegraphics[scale=1.0]{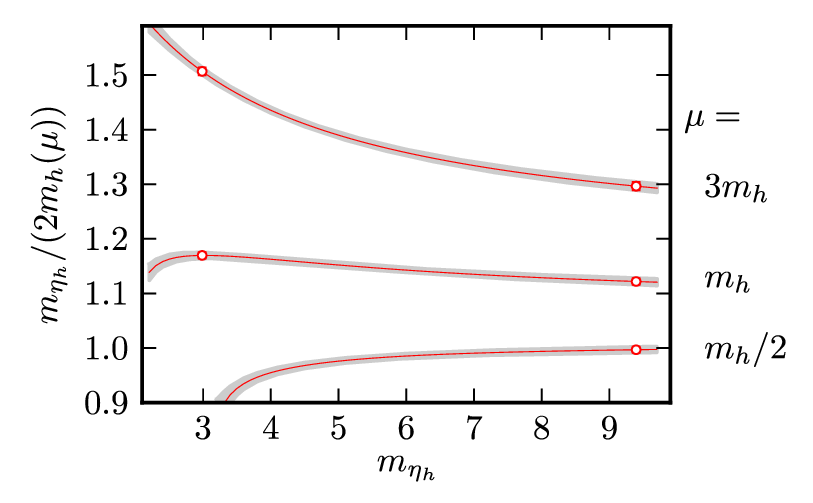}
   \caption{$z(\mu/m_h,m_{\eta_h})$ versus $m_{\eta_h}$ (in GeV)
   for three different
   values of $\mu/m_h$. The curve for $\mu\!=\!3m_h$ comes from the
   best fit to the moments. The other curves are obtained by evolving
   perturbatively from $\mu\!=\!3m_h$.}
   \label{fig:zth}
\end{figure}

\begin{figure}
   \includegraphics[scale=1.0]{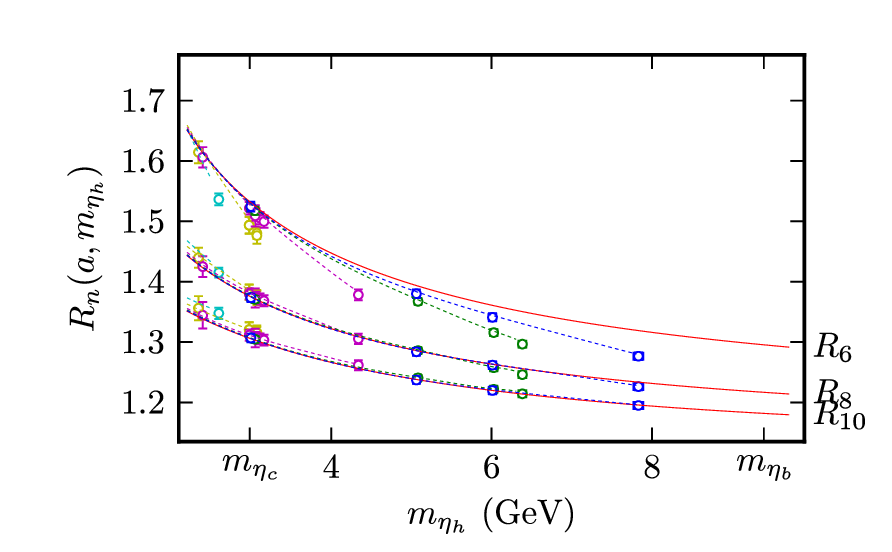}
   \caption{Simulation results for reduced moments ${R}_n$ with 
$n\!=\!6,8,10$ as functions of~$m_{\eta_h}$ for 5~different lattice 
spacings. The dashed lines show the corresponding behavior of our fit 
function, with the best-fit parameters. The curves for smaller lattice 
spacings extend further to the right. The solid lines show the 
$a\!=\!0$ limit of our best fit.}
   \label{fig:Rn}
\end{figure}

The much heavier $b$~quark is usually analyzed using effective field 
theories like NRQCD or the static-quark approximation. By using very 
small lattice spacings and the very highly improved HISQ discretization 
for the heavy quarks, we are able to extend our analysis almost to the 
$b$-quark mass, using the same relativistic discretization that we use 
for $c$~and lighter quarks. A 1.5\% extrapolation of $z(3,m_h)$, from 
the largest $m_{\eta_h}$ used in our fits to~$m_{\eta_b}$, gives us a 
new, accurate determination of the $b$-quark mass,
\begin{equation}
   m_b(10\,\mathrm{GeV},n_f=5) = \mb\,\mathrm{GeV}.
\end{equation}

This calculation demonstrates the utility of the HISQ formalism for 
studying $b$~quarks on lattices that are computationally accessible 
today. This represents a breakthrough for $b$~physics on the lattice 
since far greater precision becomes possible when all quarks are 
treated using the same formalism, and that formalism is relativistic 
and has a chiral symmetry. Even better would be to work right at the 
$b$~mass, as opposed to extrapolating from nearby; this would require a 
lattice spacing of order~$0.03$\,fm.

Both of our new $c$~and $b$~masses agree well with non-lattice 
determinations from vector-current correlators and experimental 
$e^+e^-$~collisions. A recent analysis of the continuum data 
gives~\cite{Chetyrkin:2009fv}
\begin{align}
   m_c(3\,\mathrm{GeV},n_f=4) &= 0.986(13)\,\mathrm{GeV} \\ \nonumber
   m_b(m_b,n_f=5) &= 4.163(16)\,\mathrm{GeV}
\end{align}
which compare well with our values of $\mc$\,GeV and $\mbmb$\,GeV, 
respectively. This provides strong evidence that the different 
systematic errors in each calculation are understood.

Function~$z(\mu/m_h,m_{\eta_h})$ is a by-product of our analysis. It 
relates the $\msb$~quark mass $m_h(\mu)$ to the 
$\eta_h$~mass~(\eq{z-def}). We show our result again in 
Figure~\ref{fig:zth} for $\mu\!=\!3m_h$, as well as for $\mu\!=\!m_h$ 
and $\mu\!=\!m_h/2$, which we obtain by evolving perturbatively from 
$\mu\!=\!3m_h$. The latter two curves are relatively flat, and the last 
surprisingly close to~$1$ for most masses.

Questions have been raised about the way perturbation theory is used in 
analyzing the perturbative parts of the moments~\cite{Kuhn:2010vx}. 
Like \cite{Chetyrkin:2009fv} we favor using larger scales than~$m_c$ 
for $c$-quark correlators, but, as we have shown, our results are quite 
insensitive to~$\mu$ over a broad range. Furthermore, the fact that our 
results, from pseudoscalar-density correlators, agree so well with the 
continuum results, from vector-current correlators, is also compelling 
evidence that perturbation theory is being handled correctly. We also 
find consistent results from several different moments, which is only 
possible if perturbation theory is working well. Compare, for example, 
Figure~\ref{fig:Rn} for the moments, as a function of $m_{\eta_h}$, 
with the plots of $R_n/r_n$ in Figure~\ref{fig:z}. Figure~\ref{fig:Rn} 
shows very different $m_{\eta_h}$ behavior, at the 10--20\% level, for 
different moments~$R_n$; Figure~\ref{fig:z}, where the perturbative 
part~$r_n$ is divided out, shows behavior that is almost 
moment-independent.

An additional check on our use of perturbation theory comes from the 
close agreement between our perturbative result for the ratio $m_b/m_c$ 
of $\msb$ masses (\eq{mbmc}) and our nonperturbative result for the 
ratio of HISQ masses (\eq{mbmc-nonpert}). These should be and are equal 
to within our 1\%~errors. Taken together they suggest a composite 
result of:
\begin{equation}
   \frac{m_b(\mu,n_f)}{m_c(\mu,n_f)} = 
4.51(4)\quad\quad\mbox{(composite)}.
\end{equation}

The validity of our perturbative analyses is further supported by the 
close agreement between the QCD coupling we get from the heavy-quark 
correlators,
$\alpha_\msb(M_Z)\!=\!\alphaMZ$, and that obtained from Wilson 
loops,~0.1184(6). These are radically different methods for determining 
the coupling. The first relies upon a continuum quantity, extrapolated 
to $a\!=\!0$, and continuum perturbation theory. The second relies upon 
quantities that are highly sensitive to the UV cutoff ($\pi/a$) but are 
analyzed to all orders in the cutoff using lattice perturbation theory. 
Systematic errors are almost completely different in the two cases. The 
fact that they agree to within our 0.6\%~uncertainties is highly 
nontrivial evidence that perturbative and other potential errors are 
understood.

\begin{figure}
   \includegraphics[scale=0.45]{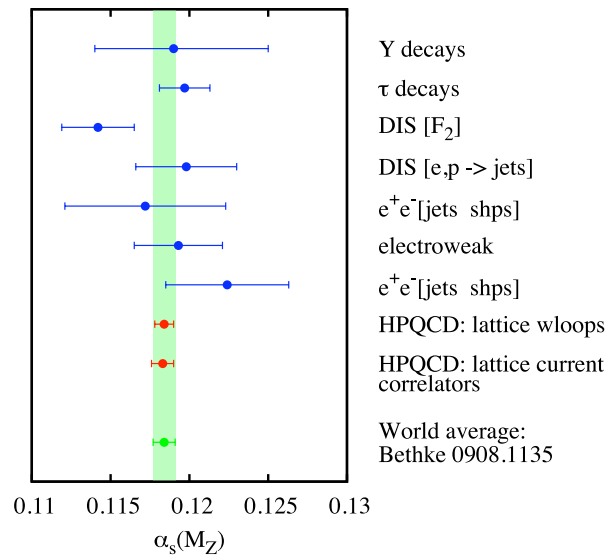}
   \caption{The 5-flavor QCD coupling $\alpha_\msb$ at the $Z$~mass as 
determined by a variety of different methods. The non-lattice numbers 
used here are from the review in~\cite{Bethke:2009jm}.}
   \label{fig:alphasummary}
\end{figure}
\begin{figure}
   \vspace{2ex}
   \includegraphics[scale=0.45]{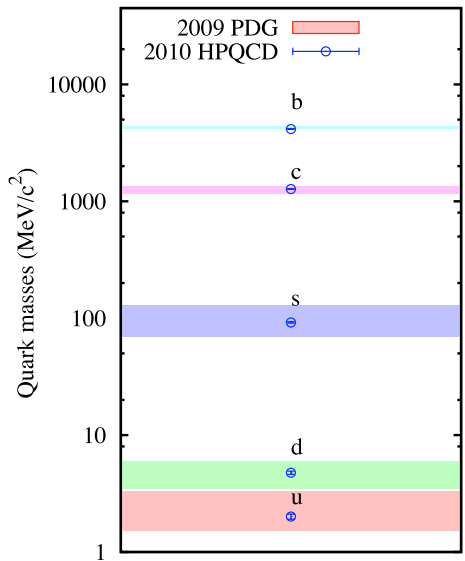}
   \caption{$\msb$ masses, for the 5~lightest quarks,
   from this paper compared with the Particle Data Group's
   current estimates~\cite{PDG}. Each mass is quoted at its 
conventional
   scale: 2\,GeV  for $u$, $d$, $s$ ($n_f\!=\!3$); $m_c$ for $c$
    ($n_f\!=\!4$); $m_b$ for $b$ ($n_f\!=\!5$).}
   \label{fig:qmass}
\end{figure}

Our coupling values also agree well with determinations
from non-lattice methods. Figure~\ref{fig:alphasummary}
summarises recent results
that were included in a world average by Bethke~\cite{Bethke:2009jm}.
The world average result, 0.1184(7), was dominated by our
previous determination from the Wilson loop analysis.
The average excluding our result was 0.1186(11), which also agrees 
well.
Including our new results into a new error-weighted world average
gives $\alpha_\msb(M_Z)\!=\!0.1184(4)$.

Our new $c$ mass is the most accurate currently available.
With it we can improve slightly on our recent determination
of light quark masses using an accurate value for $m_c/m_s$,
11.85(16), derived completely nonperturbatively from lattice
calculations~\cite{Davies:2009ih}.
Our new $c$~mass, which becomes 1.093(6)\,GeV when converted to 
$n_f\!=\!3$ at~2\,GeV, implies:
\begin{align}
m_s(2\,{\rm GeV}, n_f=3) &= 92.2(1.3)\,\mathrm{MeV}, \\ \nonumber
m_d(2\,{\rm GeV}, n_f=3) &= 4.77(15)\,\mathrm{MeV}, \\ \nonumber
m_u(2\,{\rm GeV}, n_f=3) &= 2.01(10)\,\mathrm{MeV}.
\end{align}
Our results for all 5 quark masses are compared with the Particle
Data Group's 2009 values in Figure~\ref{fig:qmass}. Agreement is 
excellent,
but our uncertainties are much smaller in every case, and by
an order of magnitude for the strange and light quarks.

Finally we note that the consistency between quark masses from lattice 
and non-lattice analyses, and between couplings from heavy-quark 
correlators and Wilson loops provides further evidence that 
taste-changing interactions in the HISQ and ASQTAD quark formalisms are 
understood and vanish as~$a\to0$. While early concerns about the 
validity of these formalisms have been largely addressed both by formal 
arguments~\cite{Sharpe:2006re,Bernard:2007eh,Kronfeld:2007ek,Golterman:2008gt,MILC,Bazavov:2009bb} and by extensive empirical studies~\cite{Follana:2006rc,Follana:2007uv,Davies:2009tsa,Davies:2009ih,Davies:2008sw,Davies:2003ik,Follana:2004sz,Follana:2005km,Follana:2005gg,Follana:2006zz}, it remains important to test the simulation technology of lattice QCD with increasing precision, given the growing importance of lattice results for phenomenology.

\section*{Acknowledgements}
We are grateful to MILC for configurations and thank Hans K\"{u}hn for 
useful discussions. Computing was done at the Ohio Supercomputer 
Centre, USQCD’s Fermilab cluster and at the Argonne Leadership 
Computing Facility supported by DOE-AC02-06CH11357. We used chroma for 
some analyses~\cite{Edwards:2004sx}. We acknowledge support by the 
STFC, SUPA, MICINN, NSF and DoE.

\newcommand{\newc}{\newcommand}
\newc{\beq}{\begin{equation}}
\newc{\eeq}{\end{equation}}
\newc{\Eq}[1]{Eq.~(\ref{#1})}
\newc{\Ref}[1]{Ref.~\cite{#1}}
\newc{\vev}[1]{\ensuremath{\left\langle {#1}\right\rangle}}
\newc{\clt}{\ensuremath{c_<}}
\newc{\cgt}{\ensuremath{c_>}}
\newc{\nlt}{\ensuremath{n_<}}
\newc{\ngt}{\ensuremath{n_>}}
\newc{\mlt}{\ensuremath{m_<}}
\newc{\mgt}{\ensuremath{m_>}}
\newc{\cnlt}{\ensuremath{c_{\nlt}}}
\newc{\cngt}{\ensuremath{c_{\ngt}}}
\newc{\cmlt}{\ensuremath{c_{\mlt}}}
\newc{\cmgt}{\ensuremath{c_{\mgt}}}

\section*{Appendix: Accelerated Fitting}
In Section~\ref{sec:finite-a} we used a trick to simplify our fits by, 
in effect, transferring fit terms from the fit function into the errors 
of the fit data. This trick can greatly speed up complicated fits. Here 
we present a formal derivation of this procedure for three increasingly 
complicated situations.

\subsection{Linear Least Squares\,---\,Exact Data}

Assuming we know $D$ values $y_i$ for a quantity $y$ which can be
expressed as a power series in $x$,
\beq
\label{eq:Series}
 y = \sum_n c_n x^n \; ,
\eeq
we wish to obtain a best fit for the first $F$ unknown coefficients 
$c_n$.
The $c_n$ are then our random variables.  If we are able to make
reasonable estimates for their means and standard
deviations $\sigma_n$, in the absence of additional information,
maximizing entropy suggests a Gaussian prior of
\beq
 P(c) \propto e^{-\sum_n c_n^2/2\sigma_n^2} \, .
\eeq
For simplicity, we assume throughout that the $c_n$ are uncorrelated 
and
have a prior mean of zero; extending to more general cases is 
straightforward.

If we knew all coefficient values, then the data $y_i$
would be completely determined, with
\beq
\label{eq:deltaDist}
 P(y|c) \propto \prod_{i=0}^{D-1} \delta(y_i - \sum_n c_n x_i^n) \, .
\eeq
Bayes' theorem
\beq
 P(c|y) \propto P(y|c) P(c)
\eeq
allows us to convert this into a distribution for $c$ given the data 
$y$.

If we are only interested in fitting a subset of coefficients $\cnlt$ 
with
$n < F$, we integrate over the remaining $\cngt$, giving
\begin{align}
P(\clt|y) &\propto e^{-\sum_{\nlt} \cnlt^2/2\sigma_n^2} \times 
\\ \nonumber
& \left[\int d\cgt \; \delta^D(y-\sum_n c_n x^n)
 \; e^{-\sum_{\ngt} \cngt^2/2\sigma_n^2}\right].
\end{align}
We replace the delta function by its Fourier representation, integrate
over first the $\cngt$, then the Fourier variables, to obtain
\begin{align}
\label{eq:LinResult}
P(\clt|y) &\propto  e^{-\sum_{\nlt} \cnlt^2/2\sigma_n^2} \times
\\ \nonumber
   &(\det \sigma_\Delta^2)^{-1/2} \;
  e^{-\Delta y\cdot(2\sigma_\Delta^2)^{-1}\cdot\Delta y}.
 \end{align}
Here
\beq
\Delta y_i \equiv y_i - \sum_{\nlt} \cnlt x_i^n
\eeq
is the discrepancy between the measured $y_i$ and the portion of the 
series
to be kept in the fit, the dot product sums over the $D$ data points,
and
\beq
 \sigma_{\Delta \, ij}^2 \equiv \sum_{\ngt}\, x^n_i\,  \sigma_n^2 \,
  x^n_j\, .
\eeq

The correlation matrix $\sigma_\Delta^2$ is independent of $\clt$ (so 
the
determinant is constant), and is the same as one would compute directly 
by
\beq
\label{eq:CorrelDefn}
 \vev{\Delta y_i \; \Delta y_j}_{\cgt} =
     \vev{\sum_{\mgt} \cmgt \; x_i^m\sum_{\ngt} \cngt \; x_j^n}_{\cgt}
\eeq
using
\beq
\vev{\cmgt \; \cngt}_{\cgt} = \sigma^2_n \, \delta_{mn} \, .
\eeq
Finally, we fit $\clt$ by minimizing $\chi^2$,
which includes these correlations and is augmented by the remaining 
$\clt$ priors.
Because the distribution is Gaussian, the $\clt$ at their minima are 
equal
to their average values.

The correlation matrix $\sigma_\Delta^2$ properly accounts for 
correlations
in the discrepancy, due to
the neglected terms, between $y$ and the portion of the series 
retained.
If $F$ terms are kept in the series, $\sigma_\Delta$ is
${\cal O}(x^F)$, enforcing agreement between $y$ and the finite series
to this order, as appropriate.
It also suggests an alternative but equivalent approach.
We may define new random (rather than exact) versions of $y$,
whose correlation matrix is $\sigma^2_\Delta$,
by moving the $\cgt$ terms to the left side of \Eq{eq:Series}.
Using the truncated series as a model for these random
data, straightforward application of Bayes' 
theorem~\cite{Lepage:2001ym}
again implies the distribution in \Eq{eq:LinResult}.

One useful consequence is that, as long as we include the correlations 
for the $\cgt$,
we may arbitrarily reduce the number of coefficients $\clt$ retained,
even to as few as one, and still obtain the same minimization values.
To see this, note that to compute a particular $\vev{\cnlt}$,
we could start with the full distribution and integrate over all $c$s. 
The
integral over $\cgt$ produces $P(\clt|y)$, which we then use in the
integral over $\clt$; the result will be the same regardless
of where the dividing line is set, as long as it does not include 
$\cnlt$.
(We could even include in $\sigma^2_\Delta$ terms of order less than 
$n$.)
Because averaging and minimization give the same result, the 
minimization
value for $\cnlt$ will also remain unchanged. This is also true of
the $c_{n_<}$~error.
While the result is
the same, reducing the number of terms in
the series to fit can significantly
improve the fitting time.

\subsection{Fits to Nonlinear Functions\,---\,Exact Data}

We now consider fitting to the data $y_i$ a general function $g_i(c_n)$ 
not
necessarily linear in the parameters $c$, and where we assume $y_i = 
g_i(c_n)$
exactly for properly chosen $c_n$.  Now
\beq
 P(y|c) \propto \prod_{i=0}^{D-1} \delta(y_i - \sum_n g_i(c_n) )\, .
\eeq
Combining with the prior $P(c)$ and integrating over the $\cgt$ gives
$P(\clt|y)$.

If our estimate of prior means is good, expanding $g$ around $\cgt = 0$
should give a reasonable approximation; an expansion
to first order gives a Gaussian.  More specifically, defining
\beq
 g_i(\clt) \equiv g_i(\clt, \cgt=0)
\eeq
and
\beq
\Delta y_i \equiv y_i - g_i(\clt) \, ,
\eeq
and integrating over $\cgt$ in this Gaussian approximation gives as 
before
\begin{align}
\label{eq:GenResult}
P(\clt|y) &\propto e^{-\sum_{\nlt} c_{n_<}^{2}/2\sigma_n^2} \times
\\ \nonumber
& (\det \sigma_\Delta^2(\clt))^{-1/2} \;
  e^{-\Delta y\cdot(2\sigma_\Delta^2(\clt))^{-1}\cdot\Delta y},
\end{align}
but with
\beq
 \sigma_\Delta^2(\clt)_{ij} \equiv \sum_{\ngt}\, \partial_n g_i(\clt)\; 
\sigma_n^2\; \partial_n g_j(\clt) \, .
\eeq
This is again the correlation one would compute directly for
$\vev{\Delta y_i\; \Delta y_j}_{\cgt}$ after expanding $g$
to first order in $\cgt$.

We have not expanded in $\clt$, so $\sigma^2_\Delta$ depends on $\clt$,
the determinant in front is not constant, and the dependence of $\Delta 
y$ on $\clt$
is not in general linear.  In practice, however, we will often further 
approximate
the distribution by setting the $\clt$ to their prior means in 
$\sigma^2_\Delta(\clt)$
before minimization.

Because $g(\clt)$ is nonlinear, $\clt$ from minimization can differ 
slightly
from $\vev{\clt}$, and due to approximations made, can vary somewhat 
with the
number of terms retained.

\subsection{Fits to Data with Intrinsic Statistical Errors}

Finally we consider the most general case, in which the data $y$ 
contribute
intrinsic statistical uncertainties in addition to those associated 
with the
truncated series.  If we measure a range of values for $y$ with an 
average
$\vev{y}$ and correlation matrix
$\sigma_y^2$, then for sufficiently large samples we expect a Gaussian 
distribution
\beq
\label{eq:yGaussian}
P(\vev{y}|c) \propto e^{-(\vev{y} - g(c))\cdot (2 \sigma^2_y)^{-1}\cdot 
(\vev{y}-g(c))}
\eeq
rather than the delta function above.  Combining with the prior $P(c)$ 
gives $P(c|\vev{y})$.

Expanding $g_i(\clt,\cgt)$ to first order around $\cgt=0$, defining
\beq
 \Delta y_i \equiv \vev{y_i} - g_i(\clt) \; ,
\eeq
and integrating $P(c|\vev{y})$ over $\cgt$ gives
\begin{align}
\label{eq:MostGenResult}
P(\clt|\vev{y}) &\propto e^{-\sum_{n_<} c_{n<}^{2}/2\sigma_n^2}  \times
\\ \nonumber
& (\det \sigma_{y\Delta}^2(\clt))^{-1/2} \;
  e^{-\Delta y\cdot(2\sigma_{y\Delta}^2(\clt))^{-1}\cdot\Delta y}.
\end{align}
The resulting correlation matrix is a combination of true statistical
and neglected series contributions, with
\beq
 \sigma_{y\Delta}^2(\clt) \equiv \sigma_y^2 + \sigma_\Delta^2(\clt) \, 
,
\eeq
as one would obtain by including both sources of uncertainty
in computing $\vev{\Delta y_i \; \Delta y_j}$ directly.  With no 
statistical
fluctuations in $y$, $\sigma_y^2=0$, and it reduces to the previous 
result.
When $\sigma_y^2$ is nonzero but small, $\sigma_\Delta^2$ still makes 
an
important contribution.

\subsection{Application to this Paper}
\begin{figure}
   \includegraphics[scale=1.0]{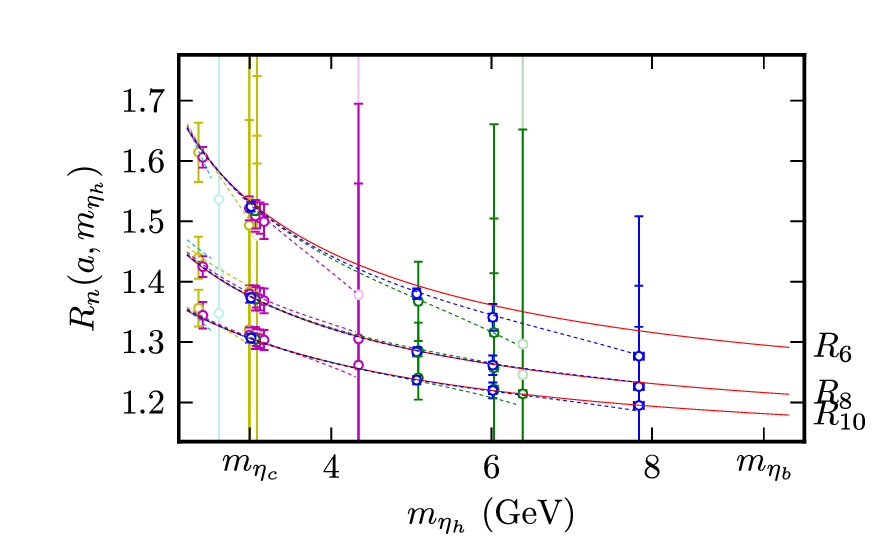}
   \caption{Same as Figure~\ref{fig:Rn} but with $N_{am}\!=\!80$ and 
$\bar{N}_{am}\!=\!3$, instead of $N_{am}\!=\!\bar{N}_{am}\!=\!30$. The 
error bars are almost entirely due to systematic errors caused by 
$am_{\eta_h}/2$ corrections omitted from the fit function.}
   \label{fig:rn-mod}
\end{figure}

We used the technique described here in much of our testing and tuning 
(but not for our final results) to speed up the 
$(am_{\eta_h}/2)^2$~fit. As described in Section~\ref{sec:finite-a}, we 
kept corrections through order $N_{am}\!=\!80$ but moved all but 
$\bar{N}_{am}\!\ll\!N_{am}$ out of the fit function and into the errors 
for the reduced moments. If we set $\bar{N}_{am}\!=\!3$, for example, 
our fit to the $R_n$~simulation data changes from Figure~\ref{fig:Rn} 
to Figure~\ref{fig:rn-mod}. The small $\bar{N}_{am}$ means that each 
point in Figure~\ref{fig:rn-mod} has much larger error bars, coming 
from $(am_{\eta_h}/2)^2$ terms moved into the~$R_n$s. The final fit 
results, however, are almost identical in both cases (to within less 
than $0.1\sigma$), with the same errors. Note that the $R_n$~errors in 
Figure~\ref{fig:rn-mod} are highly correlated, which is why the fit 
curve passes through the central value for each point. As discussed 
above these correlations are essential if results are to be independent 
of the value of~$\bar{N}_{am}$.

\bibliographystyle{plain}

\begin{thebibliography}{16}
\bibitem{mpole} We have simplified the definition of $R_n$ relative to 
our first paper by replacing the tree-level pole mass of the heavy 
quark with the bare mass parameter from the lattice. The two masses 
differ by  $\order((am_h)^4)$, which vanishes in the continuum limit.
\end{thebibliography}

\end{document}